\documentclass[se]{copernicus}
\pdfoutput=1

\usepackage[draft]{fixme}
\FXRegisterAuthor{ch}{ech}{CH}
\FXRegisterAuthor{jz}{ejz}{JZ}

\begin{document}

\linenumbers

\title{Kinematics of the South Atlantic rift}

\author[1,3]{Christian Heine}
\author[2]{Jasper Zoethout}
\author[1]{R. Dietmar M\"uller}

\affil[1]{EarthByte Group, School of Geosciences, Madsen Buildg. F09, The University of Sydney, NSW 2006, Australia}
\affil[2]{Global New Ventures, Statoil ASA, Forus, Stavanger, Norway}
\affil[3]{formerly with GET GME TSG, Statoil ASA, Oslo, Norway}

%% The [] brackets identify the author to the corresponding affiliation, 1, 2, 3, etc. should be inserted.

\runningtitle{South Atlantic rift kinematics}

\runningauthor{Heine et al.}

\correspondence{Christian Heine\\ (christian.heine@sydney.edu.au)}

\received{}
\pubdiscuss{} %% only important for two-stage journals
\revised{}
\accepted{}
\published{}

%% These dates will be inserted by the Publication Production Office during the typesetting process.

\firstpage{1}

%-----------------------------------------------------------------------------

\maketitle  %% Please note that for the copernicus2.cls this command needs to be inserted after \abstract{TEXT}

\begin{abstract}
The South Atlantic rift basin evolved as branch of a large
Jurassic-Cretaceous intraplate rift zone between the African and South
American plates during the final breakup of western Gondwana. While the
relative motions between South America and Africa for post-breakup times are
well resolved, many issues pertaining to the fit reconstruction and
particular the relation between kinematics and lithosphere dynamics during
pre-breakup remain unclear in currently published plate models. We have
compiled and assimilated data from these intraplated rifts and constructed a
revised plate kinematic model for the pre-breakup evolution of the South
Atlantic. Based on structural restoration of the conjugate South Atlantic
margins and intracontinental rift basins in Africa and South America, we
achieve a tight fit reconstruction which eliminates the need for previously
inferred large intracontinental shear zones, in particular in Patagonian
South America. By quantitatively accounting for crustal deformation in the
Central and West African rift zone, we have been able to indirectly construct
the kinematic history of the pre-breakup evolution of the conjugate West
African-Brazilian margins. Our model suggests a causal link between changes
in extension direction and velocity during continental extension and the
generation of marginal structures such as the enigmatic pre-salt sag  basin
and the S\~ao Paulo High. We model an initial E--W directed extension between
South America and Africa (fixed in present-day position) at very low
extensional velocities from 140 Ma until late Hauterivian times ($\approx$126\,Ma) when
rift activity along in the equatorial Atlantic domain started to increase
significantly. During this initial $\approx$14\,Myr-long stretching episode
the pre-salt basin width on the conjugate Brazilian and West African margins
is generated. An intermediate stage between $\approx$126\,Ma and Base Aptian is
characterised by strain localisation, rapid lithospheric weakening in the
equatorial Atlantic domain, resulting in both progressively increasing
extensional velocities as well as a significant rotation of the extension
direction to NE--SW. From base Aptian onwards diachronous lithospheric
breakup occurred along the central South Atlantic rift, first in the
Sergipe-Alagoas/Rio Muni margin segment in the northernmost South Atlantic.
Final breakup between South America and Africa occurred in the conjugate
Santos--Benguela margin segment at around 113\,Ma and in the Equatorial
Atlantic domain between the Ghanaian Ridge and the Piau\'i-Cear\'a margin at
103\,Ma. We conclude that such a multi-velocity, multi-directional rift
history exerts primary control on the evolution of this conjugate passive
margins systems and can explain the first order tectonic structures along the
South Atlantic and possibly other passive margins.
\end{abstract}

\introduction

The formation and evolution of rift basins and continental passive margins
is strongly depedendent on lithosphere rheology and strain rates
\citep[e.g.][]{Buck.PTRSLA.99, Bassi.GJI.95}. Strain rates are directly
related to the relative motions between larger, rigid lithospheric plates and
thus the rules of plate tectonics. A consistent, independent kinematic
framework for the pre-breakup deformation history of the South Atlantic rift
allows to link changes in relative plate velocities and direction between the
main lithospheric plates to events recorded at basin scale and might help to
shed light on some of the enigmatic aspects of the conjugate margin
formation in the South Atlantic, such as the pre-salt sag  basins along the
West African margin \citep[e.g.][]{Huismans.N.11, Reston.PG.10}, the extinct
Abimael ridge in the southern Santos basin, and the formation of the S\~ao
Paulo high. Over the past decades, our knowledge of passive margin evolution
and the sophistication of lithospheric deformation modelling codes has
substantially increased
\citep[e.g.][]{Huismans.N.11,PeronPinvidic.IJES.09,Ruepke.AAPG-B.08,
Crosby.EPSL.08, Lavier.Nat.06}, as have the accuracy and understanding of
global and regional relative plate motion models
\citep[e.g.][]{Seton.ESR.12,Mueller.G3.08} for oceanic areas. However, the
connections between these two scales and and the construction of quantified
plate kinematic frameworks for pre-breakup lithospheric extension remains
limited due to the fact that no equivalent of oceanic isochrons and fracture
zones are generated during continental lithospheric extension to provide
spatio-temporal constraints on the progression of extension. Provision of
such kinematic frameworks would vastly help to improve our understanding of
the spatio-temporal dynamics of continental margin formation. The South
Atlantic basin with its conjugate South American and West African margins
and associated Late Jurassic/Early Cretaceous rift structures
(Fig.~\ref{fig:rigidplates}) offers an ideal testing ground to attempt to
construct such a framework and link a kinematic model to observations from
marginal and failed rift basins. There are relatively few major lithospheric
plates involved, the motions between these plates during the early extension
phase can modeled using well-documented intraplate rifts and the conjugate
passive margins still exist in situ. In addition, an extensive body of
published literature exists which documents detailed aspects of the conjugate
passive margin architecture.

\subsection{Aims and rationale}
\label{sub:aims_and_rationale}

Following \citet{Reston.PG.10}, previously published
plate kinematic reconstructions of the South Atlantic rift have only been able
to address basic questions related to the formation of the conjugate South
Atlantic margins in pre-seafloor spreading times and fall short of
explaining the impact of plate kinematic effects on rift margin evolution. In
this paper, we present a completely revised plate kinematic model for the
pre-breakup and early seafloor spreading history of the South Atlantic rift
that integrates observations from all major South American and African rifts.
We alleviate unnecessary complexity from existing plate tectonic
models for the region -- such as a number of large, inferred intracontinental
shear zones, for which observational evidence is lacking -- while providing a
kinematic framework which can consistently explain relative plate motions
between the larger lithospheric plates and smaller tectonic blocks. The model
improves the full fit reconstruction between South American plates and
Africa, especially in the southernmost South Atlantic and is able to explain
the formation of the Salado and Colorado Basins in Argentina, two Early
Cretaceous-aged basins which strike nearly orthogonal to the main South
Atlantic rift, in the context of larger scale plate motions. This allows us
to link changes in plate motions with well-documented regional tectonic
events recorded in basins along the major rift systems and to explain the
formation of the enigmatic pre-salt sag  basins of the central South Atlantic
in the context of a multi-direction, multi-velocity plate kinematic history.

\section{Plate reconstructions: data, regional elements, and methodology} % (fold)
\label{sec:data_and_meth}

We build a spatio-temporal kinematic framework of relative motions between
rigid lithospheric plates based on the inventory of continental rifts, published
and unplublished data from the conjugate passive margins and 
information from oceanic spreading in the South Atlantic. We utilise the interactive, 
open-source plate kinematic modelling software GPlates (\url{http://www.gplates.org}) as an intergration
platform and the Generic\,Mapping Tools \citep[\url{http://gmt.soest.hawaii.edu};][]{Wessel.EOS.98} to generate our
paleo-tectonic reconstructions (See Figs.~\ref{fig:140}--\ref{fig:104} and
electronic supplements). High resolution vectorgraphics of our reconstructions
along with the plate model and geospatial data for use in GPlates are available
in the supplementary data as well as on the internet at \url{http://datahub.io/en/dataset/southatlanticrift}.

\subsection{Absolute plate motion model and timescale} % (fold)
\label{sub:timescale}

The plate kinematic model is built on a hybrid absolute plate motion model
with a moving hotspot reference frame back to 100\,Ma and an
adjusted paleomagnetic absolute reference frame for times before 100\,Ma
\citep{Torsvik.RG.08}. Global plate motion models are based on magnetic polarity timescales to
temporarily constrain plate motions using seafloor spreading magnetic
anomalies \citep[e.g.][]{Seton.ESR.12, Mueller.G3.08}. 
Here, we use the geomagnetic polarity timescale by
\citet[][hereafter called \emph{GeeK07}]{Gee.ToG.07} which places  
the young end of magnetic polarity chron M0 at 120.6\,Ma and the base of M0r at 121\,Ma. 
Agreement across various incarnations of geological time
scales exists that the base of magnetic polarity chron M0r represents the 
Barr\^emian-Aptian boundary \citep[e.g.][and references therein]{Ogg.GTS.12b, He.PEPI.08, Channell.GSAB.00}.
However, considerable debate is ongoing about the absolute age of 
of the base of M0r with proposed ages falling in two camps, one around 
121 Ma \citep[e.g.][]{He.PEPI.08, Gee.ToG.07, Berggren.95}
and a second one around 125-126 Ma \citep{Ogg.GTS.12b, Gradstein.04}. 
Both choices affect the absolute ages of the different pre-Aptian stratigraphic 
stages as magnetostratigraphy is used to provide a framework for biostratigraphic events.
Our preference of using \citet{Gee.ToG.07}'s 121.0\,Ma as base for M0r is based the arguments
put forward by \citet{He.PEPI.08} and on a review of global seafloor spreading 
velocities, in which an older age assigned to the base of M0r of 125.0\,Ma \citep{Gradstein.04} 
would result in significantly higher spreading velocities 
or larger oceanic crust flux for the pre-M0 oceanic crust \citep{Seton.G.09, Gee.ToG.07, Cogne.G3.06}. 
Additionally, \citet{Stone.JGS.08} report a reversed magnetic polarity for a Cretaceous-aged dyke 
from Pony's Pass quarry on the Falkland Islands dated to 121.3$\pm$1.2\,Ma (Ar-Ar dating). This 
further supports the notion of tying the absolute age of M0r old and Base Aptian to closer to 
121 Ma than to 125 or 126 Ma as postulated by \citet{Ogg.GTS.12a,Gradstein.04}.

% \citet[][c.f. their Fig. 5.4]{Ogg.GTS.12b} assume linearily increasing seafloor spreading
% velocities for the M-anomaly sequence of the NW Pacific, which are all tied to

\begin{figure*}[t]
\vspace*{2mm}
\begin{center}
	\includegraphics[width=12cm]{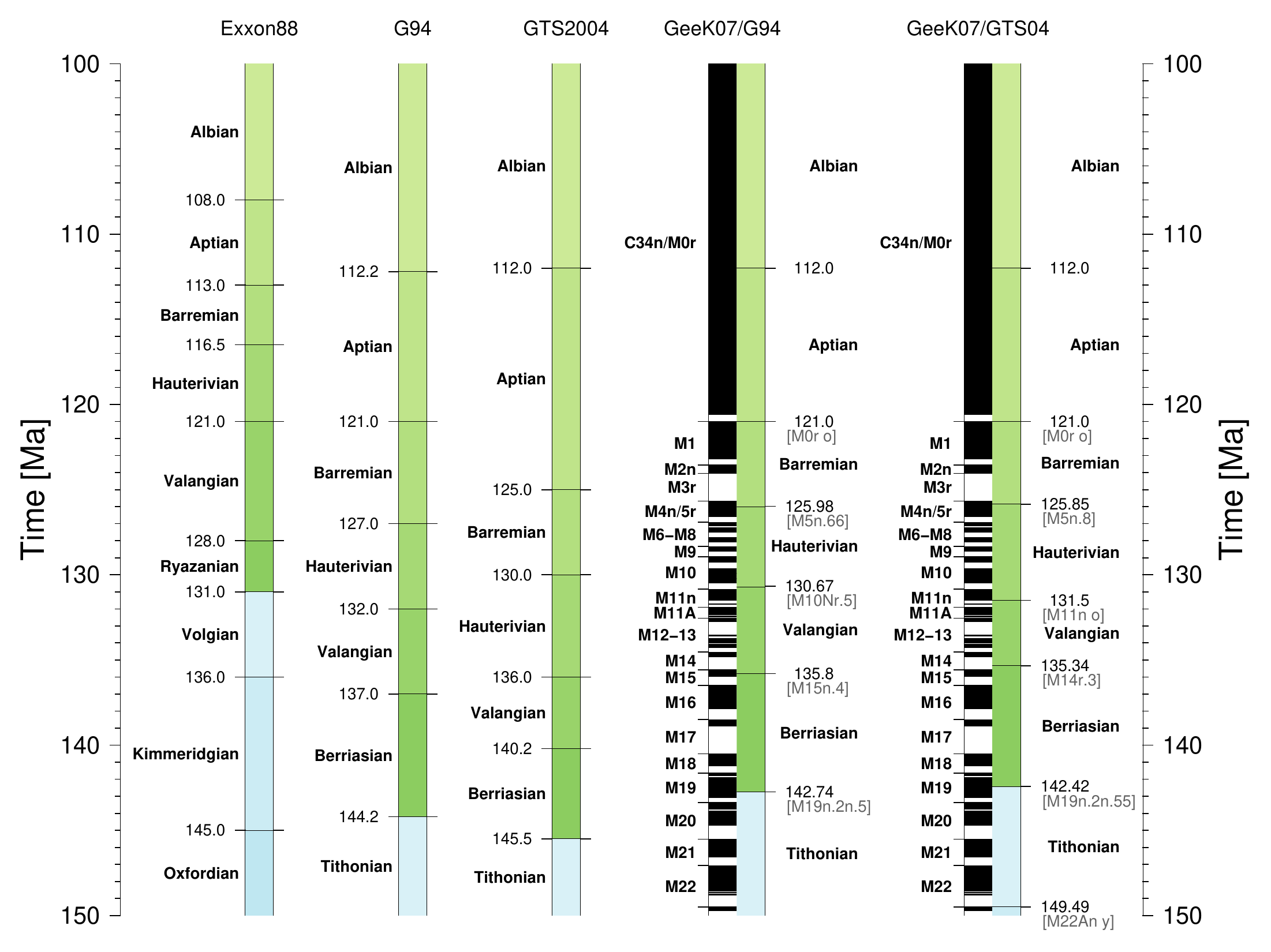}	
	\end{center}
		\caption{
		Comparison of the timescales used in publications related to the South Atlantic marginal basins and the African intraplate basins. GeeK07 is \citet{Gee.ToG.07}, Exxon88 is \citet{Haq.Sci.87}, G94 is \citet{Gradstein.JGR.94} and GTS2004 is \citet{Gradstein.04}. GeeK07+GTS04 and GeeK07+G94 shows magnetic polarity time scale with stratigraphic intervals from GTS04 and G94, respectively, adjusted to tiepoints annotated on the right hand side of the stratigraphic columns (gray font), where *.N indicates the relative position from the base of the chron (e.g. Base Barr\^emian at 125.85 ma - Base M4n young).
		\label{fig:timescalecomp}
	}
\end{figure*}

 In pre-seafloor
spreading, continental environments, well constrained direct age control
for relative plate motions does not exist. Here, stratigraphic information
from syn-rift sequences and subsidence data needs to be converted to absolute
ages in a consistent framework. As the correlation between absolute ages and
stratigraphic intervals has undergone several major iterations over the past
decades, published ages from South American and African rift basins require
readjustment. To tie stratigraphic ages to the magnetic polarity timescale
predominantly used for global plate kinematic models, we have converted the
estimates given by \citet[][here named \emph{Geek07/G94}]{Gradstein.JGR.94}
and \citet[][\emph{Geek07/GTS04}]{Gradstein.04} to the GeeK07 polarity chron
ages (Fig.~\ref{fig:timescalecomp}). We use Geek07/GTS04, which places Base
Aptian (Base M0r old) at 121\,Ma (Fig.~\ref{fig:timescalecomp}).

For published data we have converted both, stratigraphic and
numerical ages to the new hybrid timescale. A particular
example for large differences in absolute ages are the publications by
\citet{Genik.AAPG-B.93, Genik.Tectp.92} which are based on the EXXON 1988
timescale \citep[Fig.~\ref{fig:timescalecomp};][]{Haq.Sci.87}. Here, the base
Cretaceous is given as $\approx 133$\,Ma, whereas the recent GTS timescales
\citep{Gradstein.04} as well as our hybrid timescale place the base
Cretaceous between 144.2\,$\pm$\,2.5 and 142.42\,Ma, respectively. Issues 
pertaining to correlate biostratigraphic zonations across different
basins further complicate the conversion between stratigraphic and absolute
ages regionally \citep[e.g.][]{Chaboureau.Tectp.12, Poropat.GR.12}.

\subsection{Continental Intraplate Deformation}
\label{sub:subsection_name}

Deformation between rigid continental plates has to adhere to the basic rules
of plate tectonics and can be expressed as relative rotations between a
conjugate plate pair for any given time \citep{Dunbar.Tect.87}. Depending on
data coverage and quality, a hierarchical plate model can be assembled from
this information \citep[e.g.][]{Ross.Tectp.88}. In pre-seafloor spreading
environments, the quantification of horizontal deformation on regional scale
is complicated as discrete time markers such as oceanic magnetic anomalies
or clear structural features, such as oceanic fracture zones, are lacking.
However, rift infill, faulting and post rift subsidence allow -- within
reasonable bounds -- to quantify the amount of horizontal deformation
\citep[e.g.]{White.Geol.89, Gibbs.JGSL.84, LePichon.JGR.81}. 
We assume that a significant amount of horizontal
displacement ($>$15\,km) is preserved in the geological record either as
foldbelts (positive topographic features) or fault-bounded sedimentary basins
and recognised subsequently. Key elements of our plate kinematic model are a
set of lithospheric blocks which are non-deforming during the rifting and
breakup of western Gondwana (160--85\,Ma; Fig.~\ref{fig:rigidplates}). These
blocks are delineated using first order structures, such as main
basin-bounding faults, thrust belts or large gradients in reported sediment
thickness, indicative of subsurface faulting, in conjunction with potential
field and published data. The delineation of these blocks is
described in detail in later parts of this manuscript. 

We then construct a relative plate motion
model based on published data to constrain the timing,
direction, and accumulated strain in intraplate deformation zones. This
information is augmented with interpreted tectonic lineaments and kinematic
indicators (faults, strike slip zones) from various publications
\citep[e.g.][]{Matos.AGU-GM.00, Genik.AAPG-B.93, Matos.T.92a,Genik.Tectp.92, Exxon.85} 
as well as potential field data from publicly available
sources \citep{Andersen.JGeodesy.10, Sandwell.JGR.09, Maus.G3.07} to further
refine rigid block and deforming zone outlines. 
Stage rotations were derived by identifying the
predominant structural grain and choosing an appropriate rotation pole which
allows plate motions to let relative displacement occur 
so that the inferred kinematics from strain markers along the whole deforming zone (e.g. WARS/CARS) are satisfied. 
The amount of relative horizontal displacement was
then implemented by visual fitting, using published data and computed extension
estimates using total sediment thickness. 

Sediment thickness can serve as proxy for horizontal extension in 
rift basins which experienced a single phase of crustal extension. 
Here, we calculate total tectonic subsidence \citep[TTS;][]{Sawyer.JGR.85a} 
by applying an isostatic correction for varying sediment densities 
\citep{Sykes.MG.96} to the gridded total sediment thickness 
\citep{Exxon.85}. 
Extension factors based on total tectonic subsidence, assuming Airy
isostasy throughout the rifting process and not accounting for flexural rigidity, can be computed 
along a set of parallel profiles across a rift basin, oriented 
parallel to the main extension direction using the empirical 
relationship of \citet{LePichon.JGR.81}:

\begin{equation}
	TTS = 7.83 (1 - \frac{1}{\beta})
\end{equation}
\begin{equation}
	\gamma = \frac{TTS}{7.83}
\end{equation}

where $\gamma = 1 - \frac{1}{\beta}$ and $\beta$ is the stretch factor. 
It has been shown that 
this method allows to compute an upper bound for extension factors 
in rift basins, in accordance with results from other methods \citep{Barr.EPSL.85, Heine.PEPI.08}
Values obtained using this methodology are likely to 
overestimate actual extension estimates and main uncertainties in 
this approach are input sediment thickness and 
the estimated sediment compaction curves along with the assumption 
that a single rift phase created the basin. We have applied this methodology
for some of the major rift basins (Muglad/Melut, Doba/Doseo/Bongor, 
Termit/T\'en\'er\'e/Grein-Kafra, Gao, Salado, Colorado; see 
Tab.~\ref{tab:extnestimates} and supplementary material) to constrain the amount of rift-related 
displacement. We assume the bulk of the sediment thickness was accumulated
during a single rift phase. However, it is known that both CARS and WARS
experienced at least one younger phase of mild rifting and subsequent reactivation which
has affected the total sediment thickness \citep[e.g.][]{Genik.Tectp.92, Guiraud.JAfES.05} 
and hence will affect the total amount of extension computed using this method. 
The extension estimates extracted from individual rift segments
are implemented by considering the general kinematics of the whole rift
system, as individual rift segments can open obliquely and their 
margins are not necessarily oriented orthogonally to the stage pole 
small circles. 

\begin{figure}[htbp]
	\vspace*{2mm}
	\begin{center}
		\includegraphics[width=8.3cm]{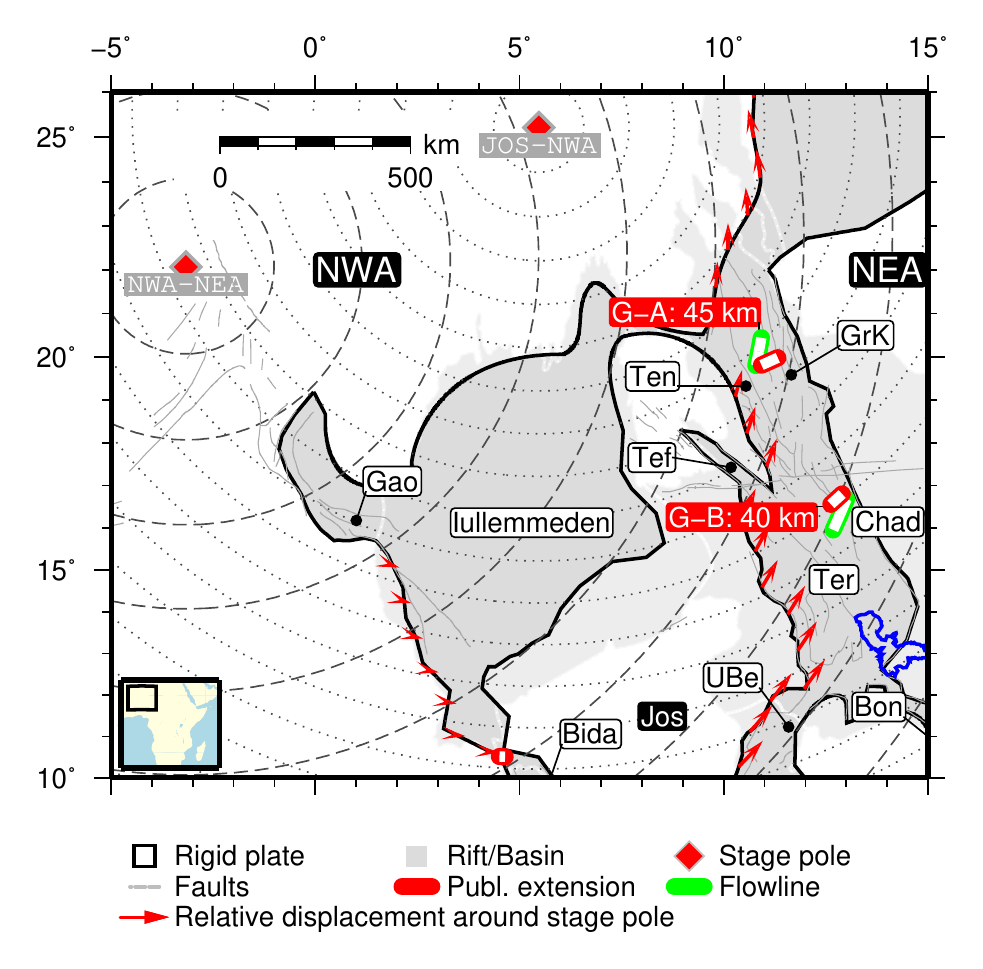}
	\end{center}
	\caption{Stage pole rotation and associated small circles (dashed and stippled grey lines) at 2$^\circ$ spacing for 145--110\,Ma interval for the West African Rift System (WARS) and associated relative plate motions at 110 ma. main plates: NEA - Northeast Africa, NWA - Nortwest Africa, Jos - Jos subplate. main rifts: Gao - Gao Trough, UBe- Upper Benoue, Ter - Termit, Bgr - Bongor, Tef - Tefidet, Ten - T\'en\'er\'e, GrK - Grein-Kafra. Basins: Chad - Chad Basin, Iullemmeden - Iullemmeden Basin, Bida - Bida Basin. Red boxes and text indicate published profiles for Grein-Kafra and Termit Rifts and estimated extension values based on \citet{Genik.Tectp.92}: G-A -- Profile A-A', G-B --  Profile B-B'.  Vectors shown on map only show relative motions, white core of published profiles and of green flowlines indicates actual amount of extension in km. These are  $\approx$45 km for profile G-A and 40--80 km for Profile G-B,  \citep[red;]{Genik.Tectp.92}, with modeled extension of $\approx$28 km and $\approx$43 km (green), respectively. Red vectors indicate angular displacement of rigid block around stage poles back to fit reconstruction position. Vectors northeast of the Grein-Kafra rift indicate possible early Cretaceous basin-and-swell topography related to strike-slip reactivation of basement structures in southern Lybia based on \citet{Guiraud.JAfES.05}.}
	\label{fig:smallcwars}
\end{figure}

Figure~\ref{fig:termit} shows  estimates for a set of
parallel profiles across the Termit Basin, orthogonal to the main 
basin axis, ranging between c.~50--100 km. Where possible, these
estimates for extension were verified using published data (Tab.~\ref{tab:extnestimates}). 
Due to the inherent lack of precise kinematic and
temporal markers during continental deformation and insufficient data we
only describe continental deformation by a single stage rotation
(Table~\ref{tab:rotationparams}, Figs.~\ref{fig:smallccars} and
\ref{fig:smallcwars}).

\subsection{Passive margins and oceanic domain}
\label{sub:passive_margins}

Over the past decade, a wealth of crustal-scale seismic data covering the
conjugate South Atlantic margins, both from industry and academic projects,
have been published \citep[e.g.][]{Blaich.JGR.11, Unternehr.PG.10, Greenroyd.GJI.07, Franke.MG.07,
Franke.GJI.06, Contrucci.GJI.04, Mohriak.GSLSP.03, Cainelli.Epis.99, Rosendahl.GSLSP.99}. We
made use of these data to redefine the location of the continent-ocean
boundary in conjunction with proprietary industry long offset reflection 
seismic data (such as the ION GXT CongoSPAN lines, 
\url{http://www.iongeo.com/Data_Library/Africa/CongoSPAN/}) as well as 
proprietary and public potential field data and models 
\citep[e.g.][]{Sandwell.JGR.09,Maus.G3.07}.

As some segments of this conjugate passive margin system show evidence for
hyperextended margins as well as for extensive volcanism and associated 
seaward dipping reflectors sequences (SDRs), we introduce the ``landward limit of the oceanic
crust'' (LaLOC) as boundary which delimits relatively homogeneous oceanic
crust oceanward from either extended continental crust or exhumed continental
lithospheric mantle landward or SDRs where an interpretation of the Moho and/or
the extent of continental crust is not
possible. This definition has proven to be useful in areas where a classic
continent-ocean boundary (COB) cannot easily be defined such as in the distal
parts of the Kwanza basin offshore Angola \citep{Unternehr.PG.10},
the oceanward boundary of the Santos basin \citep{Zalan.AAPG-ACE.11} or along the 
conjugate magmatic margins of the southern South Atlantic.

\begin{figure*}[t]
\vspace*{2mm}
\begin{center}
	\includegraphics[width=12cm]{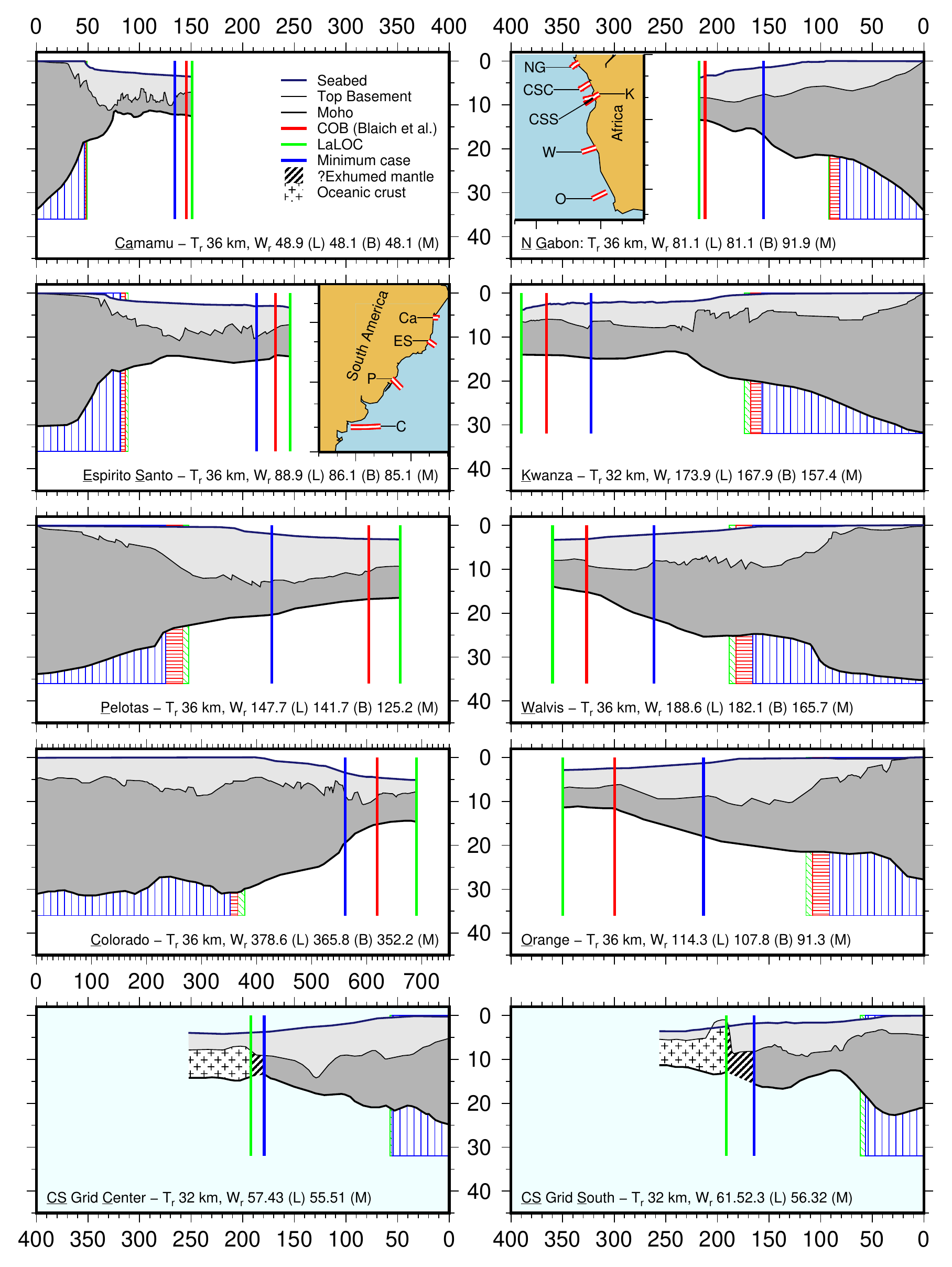}	
	\end{center}
		\caption{
		Margin cross sections and area balancing based on \citet{Blaich.GJI.09,Blaich.JGR.11} and our synthesised data based on depth-migrated and gridded CongoSPAN data along various segments of the conjugate South Atlantic margins. T$_r$: restored thickness. Extend of continental crust used for area balancing: L - LaLOC (maximum estimate, landward limit oceanic crust), M - Minimum estimate (conservative), B - COB based on \citet{Blaich.GJI.09,Blaich.JGR.11}. Note that our interpretation of the CongoSPAN data (CS Grid North and South sections) includes a zone where no Moho could be identified on seismic data, which is here tentatively interpreted as exhumed mantle. Further note the difference in length of the  Colorado Basin section. The profile data are available as plain text files at 
		\url{http://datahub.io/en/dataset/southatlanticrift}
		\label{fig:crosssections}
	}
\end{figure*}

Area balancing of the Top Basement and Moho horizons has been used on
published crustal scale passive margin cross sections by \citet{Blaich.JGR.11} to restore
the initial pre-deformation stage of the margin, assuming no out-of-plane
motions and constant area. Initial crustal thickness estimates are based on the 
CRUST2 crustal thickness model \citep{Laske.crust2.web}. 
While these assumptions simplify the actual margin
architecture and do not account for alteration of crustal thickness during
extension, Fig.~\ref{fig:crosssections} shows that the differences between a
choice of three different limits of the extent of continental crust
\citep[minimum, COB based on][ and LaLOC]{Blaich.JGR.11} only has relatively
limited effects on the width of the restored margin.

Considering the plate-scale approach of this study and inherent uncertainties
in the interpretation of subsalt structures on seismic data, 
these estimates provide valid tie points for a fit reconstruction. The
resulting fit matches well with \citet{Chang.Tectp.92}'s estimates for
pre-extensional margin geometry for the Brazilian margin (compare Figs.~\ref{fig:140}--\ref{fig:104}).
Area balancing of the continental basement (Top Basement to Moho) results in
stretching estimates ranging from 2.6-3.3 (Fig.~\ref{fig:crosssections}). 
Some of the margin cross sections do not cover the full margin width from
unstretched continental to oceanic crust (e.g. North Gabon and Orange sections).
Here, we allow for a slight overlap in the fit reconstruction. We have also 
carried out an extensive regional interpretation of Moho, Top Basement and
Base Salt reflectors on ION GXT CongoSPAN seismic data to verify results from 
areal balancing of the published data. 

The conjugate South Atlantic passive margins are predominantely non-volcanic 
in the northern and central part and volcanic south of the conjugate Santos/
Benguela segment \citep[e.g]{Blaich.JGR.11, Moulin.ESR.09}. In the volcanic
margin segments, the delineation of the extent of stretched continental crust 
is hampered by thick seaward-dipping reflector sequences (SDRs) and volcanic 
build-ups. Previous workers have associated the landward termination of the 
SDRs along the South American and southwestern African continental margins 
with the ``G'' magnetic anomaly and a prominent positive ``large magnetic
anomaly (LMA)'' delineating the boundary between a transitional crust domain
of mixed extended and heavily intruded continental crust \citep[e.g.]{
Blaich.JGR.11, Moulin.ESR.09, Gladczenko.JGSL.97, Rabinowitz.JGR.79}. We 
follow previous workers in magmatically dominated margin 
segments by using the seaward edge of SDRs and transition to normal oceanic
spreading for the location of our LaLOC. 

% \chnote{Reference to figure and implement in reconstructions}

Published stratigraphic data from the margins is integrated to constrain
the onset and dynamics of rifting along with possible extensional phases
\citep[e.g.][]{Karner.GSLSP.07b}. Little publicly available data from distal
and deeper parts of the margins exist which could further constrain 
the spatio-temporal patterns of the late synrift subsidence.

To quantify oceanic spreading and relative plate motions between the South
American and Southern African plates before the Cretaceous Normal Polarity
Superchron (CNPS, 83.5--120.6\,Ma) we use a pick database compiled by the
EarthByte Group at the University of Sydney, forming the base for the digital
ocean floor age grid \citep{Seton.ESR.12}.

We combine these data with the interpretations of \citet{Max.MPG.99} and
\citet{Moulin.ESR.09} and the WDMAM gridded magnetic data \citep{Maus.G3.07}
to create a set of isochrons for anomaly chrons M7n young (127.23 Ma), 
M4 old (126.57\,Ma), M2 old
(124.05\,Ma), and M0r young (120.6\,Ma) using the magnetic polarity
timescale of \citet{Gee.ToG.07}. M sequence anomalies from M11 to M8 are only
reported for the African side \citep{Rabinowitz.JGR.79} whereas M7 has been
identified on both conjugate abyssal plains closed to the LaLOC
\citep{Moulin.ESR.09, Rabinowitz.JGR.79}. Oceanic spreading
and relative plate velocities during the CNPS are linearily interpolated with
plate motion paths only  adjusted to follow prominent fracture zones in
the Equatorial and South Atlantic.

\subsection{Reconstruction methodology} % (fold)
\label{sub:reconstruction_methodology}

Deforming tectonic elements and their tectono-stratigraphic evolution from the South Atlantic, 
Equatorial Atlantic, and intraplate rift systems in Africa and South America 
such as intraplate basins, fault zones or passive margin segments are  
synthesised using available published and non-published data
to constructed a hierarchical tectonic model (Fig.~\ref{fig:platecircuit}). 
Starting with the African intraplate rifts, we iteratively refine 
the individual stage poles and fit reconstructions based on published estimates and our 
own computations, ensuring that the implied kinematic histories for a plate pair
do not violate geological and kinematic constraints in adajcent deforming domains. 
For example, the choice of a rotation pole and rigid plate geometries which
describe the deformation between our Southern African and
NE African plates to explain the opening of the Muglad Basin is also 
 required to match deformation along the western end of the Central African Shear
Zone in the Doba and Bongor basins (e.g.\,Fig.~\ref{fig:smallccars}).

After restoring the intra-African plate
deformation, we iteratively refine the tight fit reconstruction of South 
American plate against the NW African and South African plates by 
reconstructing our restored profiles (Fig.~\ref{fig:crosssections}), 
\citet{Chang.Tectp.92}'s restored COB, and key structural elements. 

Subsequently, remaining basin elements and rigid blocks of
the Patagonian extensional domain (Salado, North Patagonian Massif, 
Deseado, Rawson, San Julian and Falkland blocks) are restored to pre-rift/deformation stage and
integrated to achieve a full fit reconstruction for the 
latest Jurassic/Early Cretaceous time in the southern South Atlantic (cf. Fig.~\ref{fig:140}). 

% subsection reconstruction methodology (end)

\section{Tectonic elements: Rigid blocks and deforming domains}
\label{sec:tectonic_elements_rigid_blocks_and_deforming_zones}

\begin{figure*}[t]
\vspace*{2mm}
	\begin{center}
	\includegraphics[width=9cm]{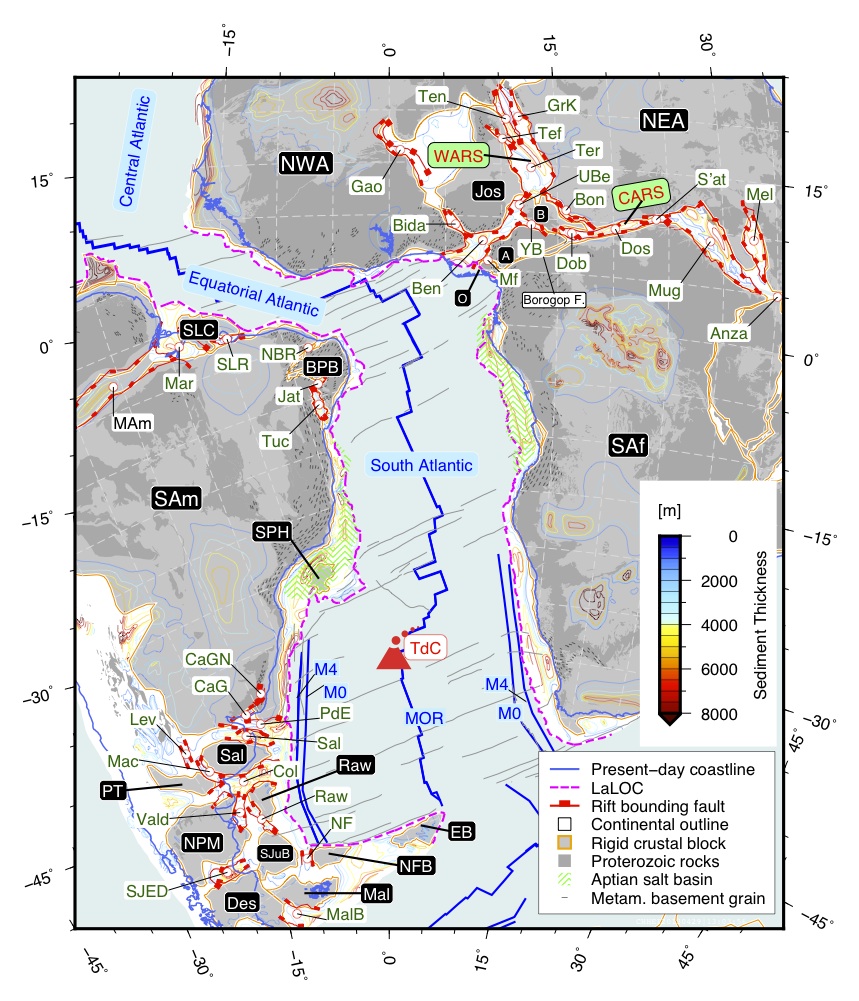}	
	\end{center}
		\caption{\label{fig:rigidplates}
		Rigid crustal blocks, Jurassic-Cretaceous rift structures, and sediment isopachs, reconstructed to 83.5\,Ma position (magnetic anomaly chron C34y) with Africa held fixed in present coordinates using the rotation poles of \citet{Nuernberg.Tectp.91}. Proterozoic rocks are based on \citet{USGS-WEP-Maps.www}, metamorphic basement grain as thin white lines \citep{Exxon.85}. Oceanic fractures zones are shown as gray lines. Tristan da Cu\~nha hotspot (TdC) indicated by volcano symbol.  Rigid crustal blocks denoted as black boxes, rift basins outlined by red fault symbols and denoted in green font and white boxes.
		Abbreviations:
		A - Adamaoua Highlands/Benoue Microplate,  
		B - Bongor Block, 
		Ben - Benoue Trough,
		BPB - Borborema Province Block,
		Bida - Bida Basin, 
		Bon - Bongor Trough,
		CaG/CaGN - Canelones Graben (Sta. Lucia)/-North, 
		Col - Colorado Rift,
		Dob - Doba Basin,
		Des - Deseado Massif in Patagonia,
		Dos - Doseo Basin,
		EWB - Maurice Ewing Bank,
		Gao - Gao Trough,
		GrK - Grein-Kafra Rift,
		Jat - Jatob\'a Rift,
		Lev - General Levalle (Leboulaye) Rift,
		Mac - Maccachin Rift, 
		MAL - Malvinas Block,
		Mal - Malvinas Basin, 
		Mar - Maranau Rift,
		Mel - Melut Rift, 
		Mf - Mamfe Basin,
		MOR - South Atlantic mid-oceanic ridge,
		Mug - Muglad Rift,
		NEA - Northeast Africa,
		NBR - Northeast Brazilian rifts,
		NF - Malvinas Norte/North Falklands,
		NFB - North Falklands Block,
		NPM - North Patagonian massif, 
		NWA - Northwest Africa,
		O - Oban Highlands Block,
		PAM - Pampean Terrane,
		PdE - Punta del Este Basin,
		RAW - Rawson Block, 
		Raw - Rawson Rift,
		SAf - Southern Africa, 
		Sal - Salado Rift,
		SAM - South America,
		SJED - San Jorge Extensional Domain,
		SJuB - San Julian Block,
		SLC - S\~ao Luis Craton,
		SLR - Sao Luis Rift 
		SPH - Sao Paulo High, 
		Tef - Tefidet Rift,
		Ten - T\'en\'er\'e Rift,
		Ter - Termit Rift,
		Tuc - Tucano Rift,
		UBe- Upper Benoue Rift,
		Yola - Yola rift branch,
		Vald - Valdes Rift.
		Sediment isopachs from \citet{Exxon.85} are shown as thin gray lines. Younger deformation and plates related rifting of East Africa Rift system are not shown. Lambert Azimuthal Equal Area projection centered at 20$^{\circ}$W/12$^{\circ}$N.
		}  
\end{figure*}

\citet{Burke.Nat.74} pointed out that Africa did not behave as a single rigid 
plate during Cretaceous rifting of 
the South Atlantic. They divided Africa into two plates separated along the 
Benue Trough-Termit Graben 
(WARS in Fig.~\ref{fig:rigidplates}). Subsequent work identified another rift 
system trending to the east from the Benoue area \citep[][ CARS in 
Fig.~\ref{fig:rigidplates}]{Genik.Tectp.92, Fairhead.Tectp.88}.

There is less evidence for Late Jurassic/Early Cretaceous deformation in South America, although several 
continent-scale strike slip zones have been postulated \citep[see][and references therein]{Moulin.ESR.09}. 
we define four major plate boundary zones and extensional domains: 
the Central African (CARS), West African (WARS), South Atlantic (SARS), 
and Equatorial Atlantic Rift Systems (EqRS). 
We include a ``Patagonian extensional domain'', composed of
Late Jurassic/Early Cretaceous aged basins and rigid blocks in southern South America in our definition of the 
SARS (Fig.~\ref{fig:rigidplates}). From these four extensional domains, 
only the SARS and EqRS transitioned from rifting to breakup, creating the 
Equatorial and South Atlantic Ocean basins. In the next sections we review timing, 
kinematics, type and amount of deformation for each of these domains. 

\subsection{Africa}
\label{sub:africa}

The West African and Central African rift systems
(Figs.~\ref{fig:rigidplates} and \ref{fig:afr}; WARS \& CARS) and associated
depocenters document extensional deformation between the following
continental lithospheric sub-plates in Africa, starting in the Latest
Jurassic/Early Cretaceous:

\begin{enumerate}
\item Northwest Africa (NWA), bound to the East by the WARS/East Niger Rift and
delimited by the Central and Equatorial Atlantic continental margins to the
West and South, respectively \citep[e.g.][]{Guiraud.JAfES.05, Burke.GSLSP.03,
Genik.Tectp.92, Fairhead.Tectp.88}.

\item Nubian/Northeast Africa (NEA), bound by the WARS to the West, and the CARS/Central
African Shear Zone to the South \citep[e.g.]{Bosworth.Tectp.92,
Genik.Tectp.92, Popoff.JAfES.88, Schull.AAPG-B.88, Browne.Tectp.85, Browne.Tectp.83}. To the East, Northeast
and North this block is delimited by the East African rift, the Red Sea and
the Mediterranean margin, respectively.

\item Southern Africa (SAf) is separated from NEA through the CARS and bound by the East
African Rift system to the East and Southeast. Its southern and western
Margins are defined by the South Atlantic continental passive margins
\citep[e.g.][]{Nuernberg.Tectp.91, Unternehr.Tectp.88}.

\item The Jos subplate, named after the Jos Plateau in Nigeria, is situated between NWA,
NEA and the Benoue Trough region. We define this plate along its western margin
by a graben system of Early Cretaceous age in the Gao Trough/Graben area in
Mali and the Bida/Nupe basin in NW Nigeria \citep{Guiraud.JAfES.05,
 Guiraud.Tectp.92a, Genik.Tectp.92, Adeniyi.PhD.84,
Cratchley.JAfES.84, Wright.Tectp.68}. The Benoue Trough and WARS delimit the
the Jos subplate to the South and East. As northern boundary we chose a diffuse
 zone through the Iullemmeden/Sokoto Basin and A\"ir massive, linking
the WARS with the Gao Trough area \citep[e.g.]{Guiraud.JAfES.05,
Genik.Tectp.92, Cratchley.JAfES.84}.

\item The Adamaoua (Benoue), Bongor  and Oban Highlands microplates 
(Fig.~\ref{fig:rigidplates}) are situated south of the Benoue Trough and north of
the sinistral Borogop fault zone. This fault zone defines the western end of
the CARS as it enters the Adamaoua region of Cameroon \citep{
Genik.Tectp.92, Benkhelil.GM.82, Burke.Nat.74}. Together with the Benoue
Trough in the north, the Atlantic margin in the west and the Doba, Bongor,
Bormu-Massenya basins it encompasses a relatively small cratonic region in
Nigeria/Cameroon which has been termed ``Benoue Subplate'' by previous
workers \citep[e.g.][]{Moulin.ESR.09, Torsvik.GJI.09}. The Yola rift branch
(YB in Fig.~\ref{fig:rigidplates}) of the Benoue Trough as well as the Mamfe Basin (Mf) indicate significant
crustal thinning \citep{Fairhead.Tectp.91b, Stuart.GJRAS.85} justifying a subdivision of this
region into the two blocks.
\end{enumerate}

The CARS and WARS are distinct from earlier Karoo-aged rift systems which
mainly affected the eastern and southern parts Africa \citep{Bumby.JAfES.05,
Catuneanu.JAfES.05} but have presumably formed along pre-existing older
tectonic lineaments of Panafrican age \citep{Daly.GSLSP.89}.

\subsubsection{Central African Rift System (CARS)}
\label{ssub:central_african_rift} 
\begin{figure*}[t]
\vspace*{2mm}
\begin{center}
	\includegraphics[width=12cm]{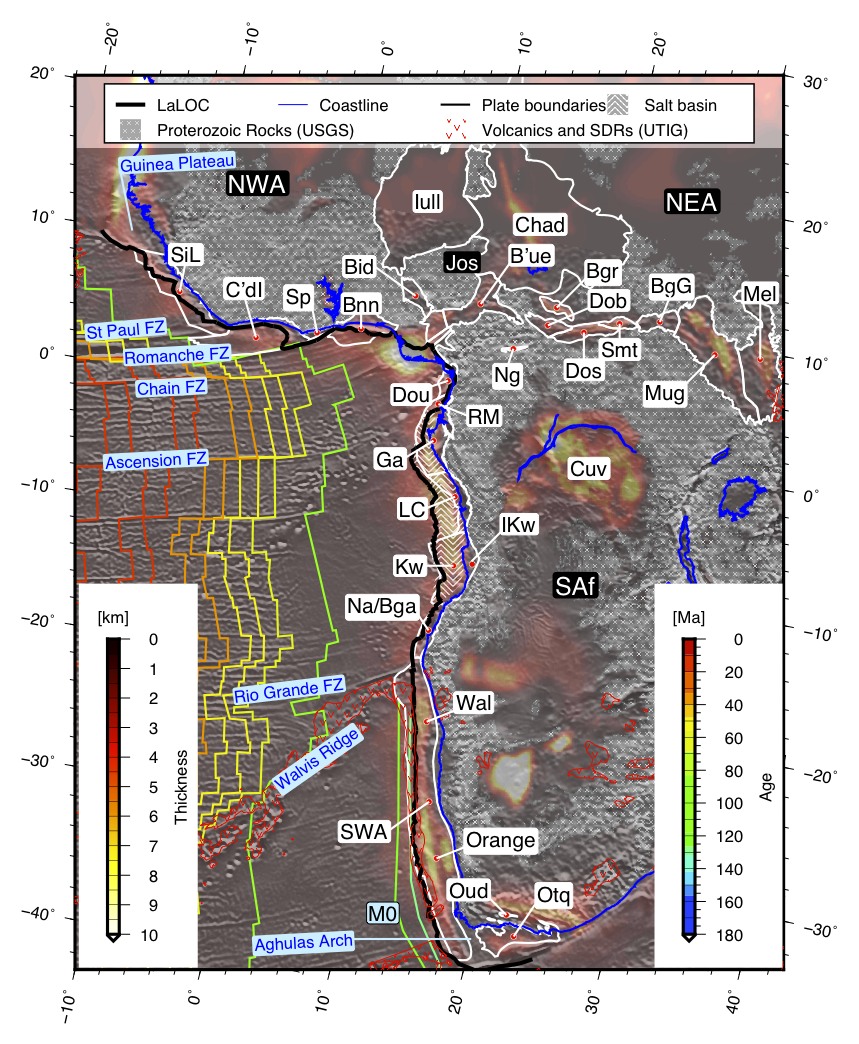}	
	\end{center}
		\caption{\label{fig:afr}
		Oblique Mercator map of present day Africa, with onshore topography \citep[ETOPO1;][]{Amante.NOAA.09}, offshore free air gravity \citep{Sandwell.JGR.09}, both grayscale, and gridded sediment thickness \citep{Exxon.85} superimposed. map legend: LaLOC - landward limit of oceanic crust, Proterozoic rocks and Mesozoic extrusive rocks based on \citet{USGS-WEP-Maps.www}.  Seaward-dipping reflectors (SDRs) based on UTIG PLATES open data (\url{https://www.ig.utexas.edu/research/projects/plates/data.htm}). White polygons indicate sedimentary basins (N to S): SiL - Sierra Leone marginal, C'dI - C\^ote d'Ivorie marginal, Sp - Saltpond, Bnn - Benin, Bid - Bida, Iull - Iullememmden, Chad - Chad Basin, B'ue - Benoue Trough, Bgr - Bongor Trough, Dob - Doba, Dos - Doseo, BgG - Banggara Graben, Mel - Melut, Mug - Muglad, Smt - Salamat, Ng - Ngaoundere, Dou - Douala, RM - Rio Muni, Ga - Gabon Coastal, Cuv - Cuvette Central, LC - Lower Congo, Kw - Kwanza/Cuanza, IKw - Inner Kwanza, Na/Bga - Namibe/Benguela, Wal - Walvis, SWA - South West African marginal, Orange - Orange, Oud - Oudtshoorn, Otq - Outeniqua. Fz: Oceanic fracture zone. Colored lines are oceanic isochrons from \citep{Mueller.G3.08}, M0 - M0 magnetic anomaly (120.6 ma). Plate and sub-plates: NWA - Northwest Africa, NUB--Nubia/Northeast Africa, JOS- Jos Plateau Subplate and SAf -- Austral Africa. Origin of map projection at 10$^{\circ}S$ 5$^{\circ}$W, azimuth of oblique equator 80$^{\circ}$.
		}
\end{figure*}

The eastern part of the CARS, consisting of
the Sudanese Melut, Muglad and Bagarra basins, forms a zone a few hundred
kilometers wide with localised NW-SE striking sedimentary basins which are
sharply delimited to the north by the so-called Central African Shear Zone
\citep{Bosworth.Tectp.92, McHargue.Tectp.92, Schull.AAPG-B.88, Exxon.85,
Browne.Tectp.83}. Subsurface structures indicate NNW/NW-trending main
basin-bounding lineaments and crustal thinning with up to 13\,km of Late
Jurassic/Early Cretaceous-Tertiary sediments
\citep[Fig.~\ref{fig:afr};][]{Mohamed.JAfES.01,
McHargue.Tectp.92, Schull.AAPG-B.88, Browne.Tectp.85, Browne.Tectp.83}. Published values for
crustal extension in the Muglad and Melut basins in Sudan range between 22-48\,km \citep{Browne.Tectp.83},
15--27\,km \citep{McHargue.Tectp.92} in SW-NE direction and $\beta = 1.61$
for an initial crustal thickness of 35\,km, resulting in 56\,km of extension
\citep{Mohamed.JAfES.01} with a first postrift phase commencing in the Albian
\citep[$\approx$110\,Ma;][]{McHargue.Tectp.92}.

Dextral transtensional motions along the western part of the CARS during the
Early Cretaceous created the Salamat, Doseo and Doba basins
\citep[Fig.~\ref{fig:afr};][]{Bosworth.Tectp.92,
McHargue.Tectp.92, Schull.AAPG-B.88}. These depocenters locally contain more than 8\,km of
Early Cretaceous to Tertiary sediments and are bound by steeply dipping
faults, indicative of pull-apart/transtensional kinematics for the basin
opening \citep[Fig.~\ref{fig:afr};][]{Genik.AAPG-B.93,Maurin.Tectp.93,
Binks.Tectp.92, Genik.Tectp.92, Guiraud.Tectp.92b, Exxon.85,
Browne.Tectp.83}. The structural inventory of these basins largely follows
old Panafrican-aged lineaments and has been reactived during the Santonian
compressive event \citep{Guiraud.JAfES.05, Janssen.GSA-B.95, Maurin.Tectp.93,
Genik.Tectp.92, Daly.GSLSP.89}.

\begin{figure}[htbp]
	\vspace*{2mm}
	\begin{center}
		\includegraphics[width=8.3cm]{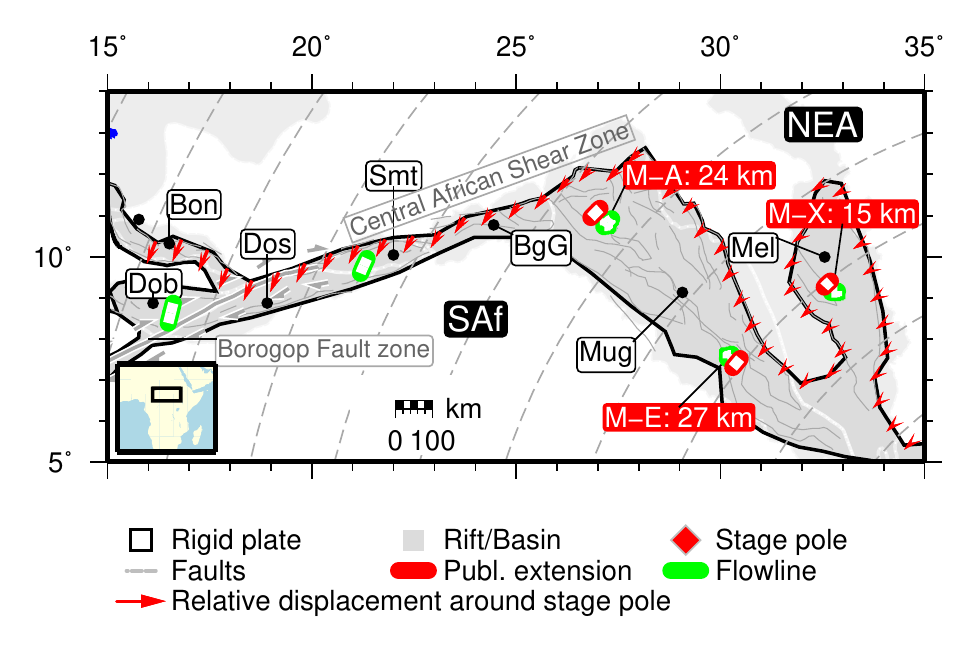}
	\end{center}
	\caption{Stage pole rotation and associated small circles (dashed grey lines) at 2$^\circ$ spacing for 145--110 ma interval for the Central African Rift System (CARS) and associated relative plate motions at 110\,Ma. main rigid plates: SAf - Southern Africa, NEA - Northeastern Africa. main basins and rifts: Bgr - Bongor Basin, Dob - Doba, Dos - Doseo, Smt - Salamat, BgG - Bangara Graben, Mug- Muglad, Mel - Melut, Anza - Anza Rift. Red boxes and text indicate published profiles for Melut and Muglad Basins and estimated extension values based on \citet{McHargue.Tectp.92}: M-A -- Profile A-A', M-E --  Profile E-E', M-X -- Profile X-X'. Vectors shown on map only show relative motions, white core of published profiles and of green flowlines indicates actual amount of extension in km. These are $\approx$24 km for profile M-A, $\approx$27 km for profile M-E and $\approx$15 km for Profile M-X during the intial basin formin phase F1\citep[red;]{McHargue.Tectp.92}, with modeled extension of $\approx$30 km,  $\approx$17 km, and $\approx$17 km (green), respectively. Red vectors indicate angular displacement of rigid block around stage pole back to fit reconstruction position. }
	\label{fig:smallccars}
\end{figure}

The Borogop Fault (Fig.~\ref{fig:rigidplates} \& \ref{fig:smallccars}) defines the western 
part of the Central African Shear Zone and enters the cratonic area of the Adamaoua uplift in Cameroon, where
smaller, Early Cretaceous-aged basins such as the Ngaoundere Rift are located
\citep[Fig.~\ref{fig:afr};][]{Maurin.Tectp.93, Plomerova.GJI.93, Guiraud.Tectp.92b, Fairhead.Tectp.91}. 
The reported total dextral displacement is estimated to be 40--50\,km based on basement outcrops and around 35 km in the Doseo Basin \citep{Genik.Tectp.92, Daly.GSLSP.89}. 

% FIXME: write about the requirements and setup in the CARS that the stage pole needs to satisfy multiple observations.

We have implemented a compounded total extension between 20--45\,km between 140 and 110\,Ma 
(early Albian) in the Muglad and Melut Basins between, generated through rotation of
the SAf block counterclockwise relative to NEA (Fig.~\ref{fig:smallccars}, 
Tab.~\ref{tab:extnestimates}). The stage pole is located located in 
the Somali Basin, south of the Anza Rift/Lamu Embayment, resulting in
 moderate extension ($\approx$10\,km) in the northern Anza Rift in Kenya which 
is supported by observations faulting of Early Cretaceous age \citep{Morley.AAPG-SiG.99, Reeves.EPSL.87}. 
The chosen stage rotation results in estimates of distributed oblique extension/
sinistral transtension along the western 
end of the CARS in the Bongor, Doba and Doseo Basins. Based on our choice of rigid plates,
this stage pole accounts for the opening of the Sudanese, Central African Rifts as well as the Bongor
basin. 

Transtensional dextral displacement along the Borogop fault zone in our model amounts
 to 40-45 km in the Doseo Basin which is in agreement with 25--56\,km extension reported
from the Sudanese basins \citep{Mohamed.JAfES.01, McHargue.Tectp.92}. 

Stage poles 
and associated small circles for the NEA--SAf rotation are oriented orthogonally to 
mapped Early Cretaceous extensional fault trends for the Doba and Doseo Basin \citep{Genik.Tectp.92},
and graben-bounding normal faults in the Sudanese basins
(Fig.~\ref{fig:smallccars}). Other authors have used 70\,km of strike
slip/extension for the CASZ and Sudan Basins, respectively
\citep{Moulin.ESR.09}, which is about double the amount reported
\citep{Genik.Tectp.92, McHargue.Tectp.92}. \citet{Torsvik.GJI.09} model the
CARS but do not specify an exact amount of displacement between their
Southern African and the NE African sub-plates.

\subsubsection{West African Rift System (WARS)}
\label{ssub:west_african_rift}

\begin{figure}[htbp]
	\vspace*{2mm}
	\begin{center}
		\includegraphics[width=8.0cm]{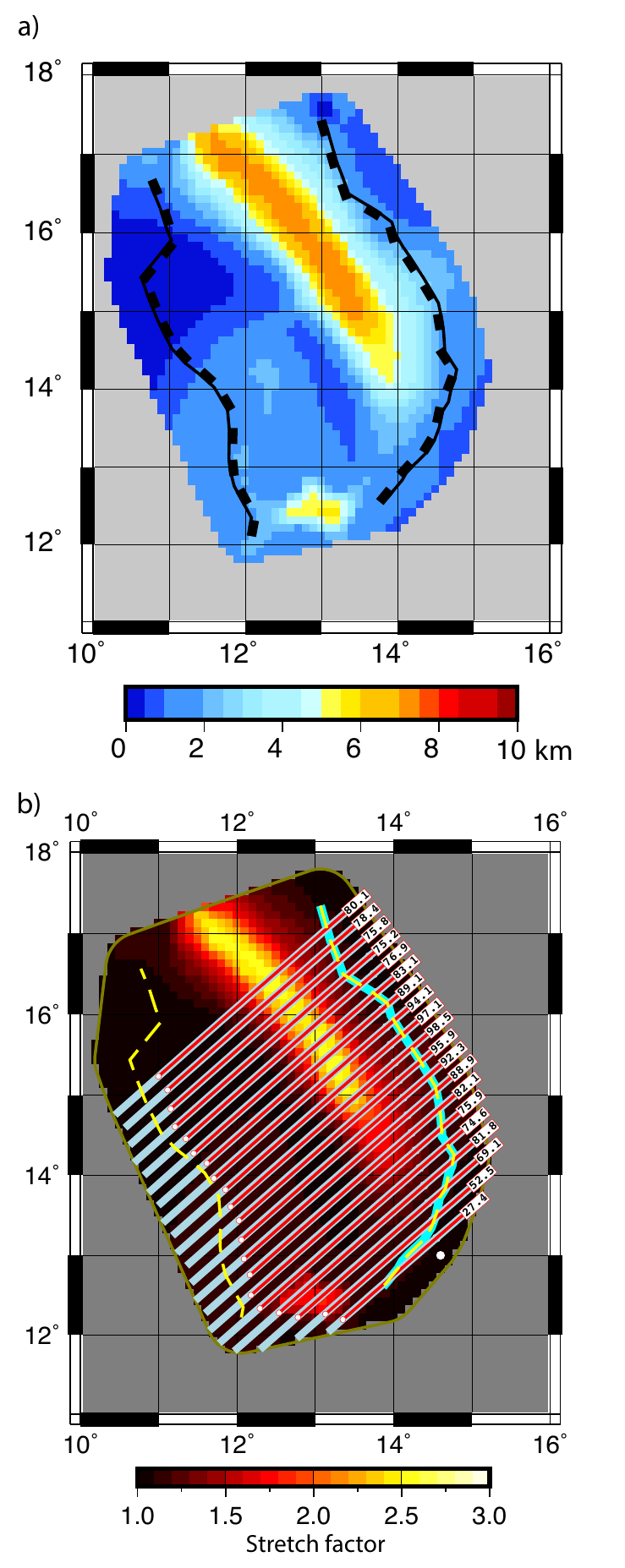}
	\end{center}
	\caption{Computed extension for the Termit Basin in Chad based on total sediment thickness. a) sediment thickness based on \citet{Exxon.85}; b) stretching factor grid and computed amount of extension using the method of \citet{LePichon.JGR.81}, see text for details. Estimates of horizontal extension for the Termit basin. Cyan colored lines are main rift bounding faults, light blue lines are profile lines with original length, red lines are restored length based on extension estimates from sediment thickness. Number indicate individual amount of computed extension along each line in kilometers.} 
	\label{fig:termit}
\end{figure}

The West African/East Niger rift (WARS) extends northward from the eastern
Benoue Trough region through Chad and Niger towards southern Algeria and
Lybia (Fig.~\ref{fig:rigidplates}). The recent Chad basin is underlain by a
series of N--S trending rift basins, encompassing the Termit Trough, N'Dgel
Edgi, Tefidet, T\'en\'er\'e, and Grein-Kafra Basins containing up to 12\,km
of Early Cretaceous to Tertiary sediments
\citep[Figs.~\ref{fig:afr} \& \ref{fig:termit};][]{Guiraud.JAfES.05, Guiraud.Tectp.92a,
Genik.Tectp.92, Exxon.85}. These basins are extensional, asymmetric rifts,
initiated through block faulting in the Early Cretaceous, with a dextral
strike-slip component reported from the Tefidet region \citep[``Tef'' in
Fig.~\ref{fig:rigidplates};][]{Guiraud.Tectp.92a, Genik.Tectp.92}. The infill
is minor Paleozoic to Jurassic pre-rift, non-marine sediments and a
succession of non-marine to marine Cretacous clastics of up to 6\,km
thickness with reactivation of the structures during the Santonian
\citep{Bumby.JAfES.05, Guiraud.JAfES.05, Genik.Tectp.92}. Towards the north
of the WARS, N--S striking fault zones of the El Biod-Gassi Touil High in the
Algerian Sahara and associated sediments indicate sinistral transpression
during the Early Cretaceous \citep{Guiraud.Tectp.92a}. The main rift
development occurred during \citet{Genik.Tectp.92}'s Phase 3 from the Early
Cretaceous to Top Albian (130--98\,Ma) with full rift development by 108\,Ma
\citep[][using the EXXON timescale]{Genik.Tectp.92}.

Early Cretacous sedimentation and normal faulting in the Iullemmeden/Sokoto
and Bida Basins in NW Nigeria and the Gao Trough in mali indicates that
lithospheric extension also affected an area NW of the Jos subplate and
further west of the WARS sensu strictu \citep{Guiraud.JAfES.05,
Obaje.AAPG-B.04, Genik.AAPG-B.93, Genik.Tectp.92, Guiraud.Tectp.92a, 
Adeniyi.PhD.84, Cratchley.JAfES.84, Petters.EGH.81, Wright.Tectp.68}.
Reported sediment thicknesses here range between 3--3.5\,km for the Bida
Basin \citep[][Fig.~\ref{fig:afr}]{Obaje.AAPG-B.04}. Our definition of the
WARS hence encompasses this area of diffuse lithospheric extension.

Palinspastic restoration of 2-D seismic profiles across the Termit basin part
of the WARS yields extension estimates between 40--80\,km based \citep[using
Moho depths of 26\,km;][]{Genik.Tectp.92}. Our computed maximum extension
estimates for the WARS rift basins using total tectonic subsidences results in 
a maximum cumulative extension of 90--100\,km (Fig.~\ref{fig:termit}, Tab.~\ref{tab:extnestimates}). 
We
use 70\,km of extension in the Termit Basin region and 60\,km in the
Grein-Kafra Basin to accommodate relative motions between the Jos Subplate
and NEA between Base Cretaceous and 110\,Ma. Fault and sediment isopach
trends indicate an E--W to slightly oblique rifting, trending NNW-SSE for the
main branch of the WARS. Our stage rotation between NWA and NEA results
in oblique, NNE-SSW directed opening of the WARS (Fig.~\ref{fig:smallcwars}).
Other workers have used 130\,km of E--W directed extension in the South
(Termit Basin) and 75\,km for northern parts \citep[Grein/Kafra
Basins;][]{Torsvik.GJI.09} between 132--84\,Ma or 80\,km of SW-NE directed
oblique extension \citep{Moulin.ESR.09}. For the Bida (Nupe) Basin/Gao
Trough, we estimate that approximately 15\,km of extension occurred between
the Jos Subplate and NWA, resulting in a cumulative extension between NWA and
NEA of 85--75\,km between 143\,Ma and 110\,Ma.

\subsubsection{Benoue Trough}
\label{ssub:benoue_trough} The Benoue Trough and associated basins like the
Gongola Trough, Bornu and Yola Basins are located in the convergence of the
WARS and CARS in the junction between the Northwest, Northeast and Southern
African plates. The tectonic position makes the Benoue Trough susceptible to
changes in the regional stress field, reflected by a complex structural
inventory \citep{Benkhelil.JAfES.89, Popoff.JAfES.88}. Sediment thicknesses
reach locally more than 10\,km along, with the oldest outcropping sediments
reported as Albian age from anticlines in the Upper Benoue Trough
\citep{Fairhead.JAfES.90, Benkhelil.JAfES.89}. Subsidence in the Benoue
Trough commences during Late Jurassic to Barr\^emian as documented by the
Bima-1 formation in the Upper Benoue Trough \citep{Guiraud.JAfES.05,
Guiraud.Tectp.92a}.

The observed sinistral transtension in the Benoue Trough is linked to the
opening of the South Atlantic basin and extension in the WARS and CARS
\citep{Guiraud.JAfES.05, Genik.AAPG-B.93, Genik.Tectp.92, Fairhead.Tectp.91,
Fairhead.JAfES.90, Benkhelil.JAfES.89, Popoff.JAfES.88, Benkhelil.GM.82,
Burke.Tectp.76, Burke.Nat.74}. It follows that the onset of rifting and
amount of extension in the Benoue Trough is largely controlled by the
relative motions along the WARS and CARS. maximum crustal extension estimates
based in gravity inversion are 95\,km, 65\,km, and 55\,km in the Benue and
Gongola Troughs, and Yola Rift with about 60\,km of sinistral strike slip
\citep{Fairhead.JAfES.90, Benkhelil.JAfES.89}. Rift activity is reported from
the ``Aptian (or earlier)'' to the Santonian \citep{Fairhead.JAfES.90},
synchronous with the evolution of the CARS and WARS \citep{Guiraud.JAfES.05,
Genik.Tectp.92}.

We here regard the Benoue Trough as product of differential motions between
NWA, NEA and the Adamaoua Microplate which is separated from South Africa by the Borogop Fault zone. 
Deformation implemented in our model for the WARS and CARS
result in $\approx$20\,km of N--S directed relative extension and about
50\,km of sinistral strike-slip in the Benoue Trough between the Early
Cretaceous and early Albian.

The Mamfe Basin at the SW end of the Benoue Trough is separating the Oban Highlands Block
from the Adamaoua Microplate in the east (Fig.~\ref{fig:rigidplates}). It is conjugate to NE Brazil. 20 km of extension are reported
for this region which we have restored in our plate model.

\subsection{South America}
\label{sub:overview_south_america}

\begin{figure*}[t]
\vspace*{2mm}
\begin{center}
	\includegraphics[width=12cm]{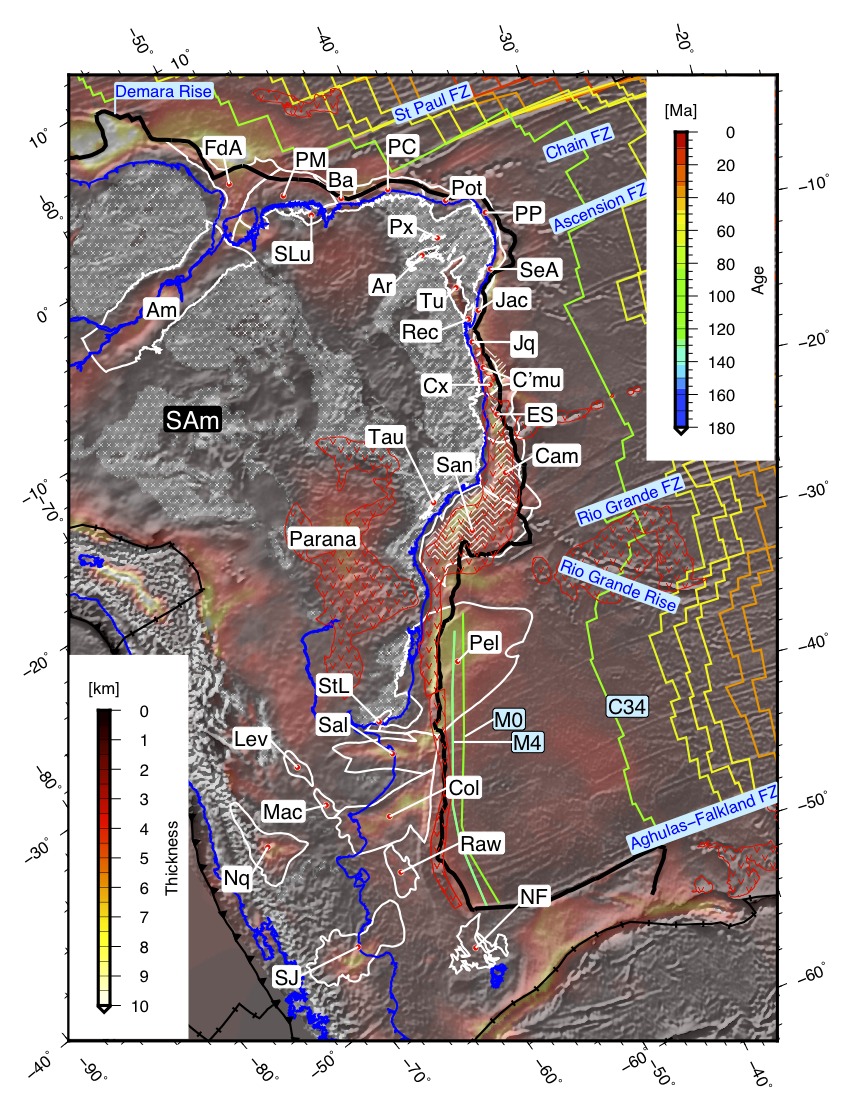}
	\end{center}
		\caption{\label{fig:sam}
		Oblique Mercator map of present day South America (SAM), with onshore topography \citep[ETOPO1;][]{Amante.NOAA.09}, offshore free air gravity \citep{Sandwell.JGR.09}, both grayscale, and gridded sediment thickness \citep{Exxon.85} superimposed. map legend as in Fig.~\ref{fig:afr}. White polygons denote sedimentary basins (N to S): FdA - Foz do Amazon, PM - Para-Maranhao, Ba - Barreirinhas, PC - Piaui-Ceara, SLu - Sao Luis Graben, Px - Rio do Peixe, Pot - Potiguar, PP - Pernambuco--Paraiba, Ar - Araripe, SeA - Sergipe-Alagoas, Tu - Tucano, Re - Rec\^oncavo, Jac - Jacuipe, Am - Amazonas, C'mu - Camamu, Cx - Cumuruxatiba, ES - Espirito Santo, Cam - Campos, San - Santos, Tau - Taubat\'e, Parana - Parana Basin, Pel - Pelotas, StL - Santa Lucia, Sal - Salado, Lev - General Levalle, Col - Colorado, mac - maccachin, Nq - Neuq\'en, Raw - Rawson, NF - North Falklands, SJ - San Jorge
		Colored lines are oceanic crust ages from \citep{Mueller.G3.08}. Plate boundaries (thin back decorated lines) after \citet{Bird.G3.03}.  Origin of map projection at 50$^{\circ}S$ 20$^{\circ}$W, azimuth of oblique equator -65$^{\circ}$.
		}
\end{figure*}

The present-day South American continent is composed of a set of Archean and
Proterozoic cores which were assembled until the early Paleozoic, with its
southernmost extent defined by the Rio de la Plata craton
\citep[Fig.~\ref{fig:sam};][]{Pangaro.MPG.12, Almeida.ESR.00}. Large parts of
South America, in contrast to Africa, show little evidence for significant
and well preserved, large offset (10's of km) intraplate crustal deformation
during the Late Jurassic to Mid-Cretaceous. In the region extending from the
Guyana shield region in the north, through Amazonia and S\~ao Francisco down
to the Rio de la Plata Craton there are no clearly identifiable sedimentary
basins or compressional structures with significant deformation reported in
the literature which initiated or became reactivated during this time
interval.

The Amazon basin and the Transbrasiliano Lineament have been used as the two
major structural elements by various authors to accommodate intraplate
deformation of the main South American plate. \citet{Eagles.GJI.07} suggests
the Solim\~oes-Amazon-Maraj\'o basins as location of a temporary,
transpressional plate boundary during South and Equatorial Atlantic rifting,
where a southern South America block is dextrally displaced by
$\approx$200\,km against a northern block. The basin is underlain by old
lithosphere of the Amazonia Craton \citep{Li.PR.08, Almeida.ESR.00} which
experienced one main rifting phase in early Paleozoic times and subsequent,
predominantely Paleozoic sedimentary infill
\citep[Fig.~\ref{fig:sam};][]{
CruzCunha.BGP.07, Gonzaga.AAPG-M.00, Matos.T.92b, Nunn.JGR.88}. It is covered by a thin blanket of Mesozoic and Cenozoic
sediments which show mild reactivation with NE-trending reverse faults and
minor dextral wrenching along its eastern margin/Foz do Amazon/Maraj\'o Basin
during the Late Jurassic to Early Cretaceous \citep{CruzCunha.BGP.07,
Costa.JSAES.01, Gonzaga.AAPG-M.00}. Reactivation affecting the whole Amazon
basin is reported only from the Cenozoic \citep{Azevedo.PhD.91, Costa.JSAES.01}. We do not regard this tectonic element as a temporary plate
boundary during formation of the South Atlantic.

The continental-scale Transbrasiliano lineament
\citep[TBL;][]{Almeida.ESR.00} formed during the Pan-African/Brasiliano
orogenic cycle. It is a potential candidate for a major accommodation zone
for intraplate deformation in South America \citep{Feng.JGR.07, PerezGussinye.G3.07}, however, the amount of accommodated deformation and the exact
timing remain elusive \citep{Almeida.ESR.00}. Some authors suggest strike
slip motion along during the opening of the South Atlantic between
60--100\,km along this 3000\,km long, continent-wide shear zone, reaching
from NE Brazil down into northern Argentina \citep{Aslanian.GJI.10,
Moulin.ESR.09,Fairhead.SEG.07}. Along undulating lineaments such as the TBL,
any strike-slip motion would have resulted in a succession of restraining and
releasing bends \citep{Mann.GSLSP.07a}, creating either compressional or
extensional structures in the geological record. For comparison, the reported
offset along the Borogop Shear zone in the Central African Rift System ranges
around 40\,km during the Late Jurassic--Early Cretaceous and created a series
of deep ($>$6\,km) intracontinental basins \citep[Doba, Doseo, Salamat --
Fig.~\ref{fig:afr}; ][]{Genik.Tectp.92, McHargue.Tectp.92}. While we do not
refute evidence of reactivation of the TBL during the opening of the South
Atlantic, published geological and geophysical data do not provide convincing
support for the existence of a plate boundary separating the South American
platform along the Transbrasiliano lineament during the opening of the South
Atlantic.

In southern Brazil, previous authors have argued for the Carboniferous
Paran\'a Basin being the location for a large NW--SE striking
intracontinental shear zone to close the ``underfit'' problems in the
southern part of the South Atlantic \citep{Moulin.GSLSP.12, Moulin.ESR.09,
Torsvik.GJI.09, Eagles.GJI.07, Nuernberg.Tectp.91, Unternehr.Tectp.88,
Sibuet.DSDP.84}. This zone, obscured by one of the largest continental flood
basalt provinces in the world \citep{Peate.AGU-GM.97, White.JGR.89}, has been
characterised as R-R-R triple junction with 100\,km N-S extension
\citep{Sibuet.DSDP.84}, as Paran\'a-Coehabamba shear zone with 150\,km
dextral offset \citep{Unternehr.Tectp.88}, as Parana-Chacos Deformation Zone
with 60--70\,km extension and 20-30\,km of strike slip
\citep{Nuernberg.Tectp.91}, as Paran\'a-Etendeka Fracture Zone -- a
transtensional boundary with 175\,km lateral offset \citep[]{Torsvik.GJI.09},
or dextral strike-slip zone with 150 km strike slip and 70\,km extension
\citep{Moulin.ESR.09}. \citet{Peate.AGU-GM.97} ruled out the possibility for
a R-R-R triple junction due to timing of magma emplacement and orientation
of the associated dyke swarm. The minimum amount of deformation proposed by
previous authors is 20--30\,km strike slip and 60--70\,km extension
\citep{Nuernberg.Tectp.91}. In analogy to the well documented CARS and the
discussion of the TBL above, such significant displacement should have
resulted in a set of prominent basins extending well beyond the cover of the
Paran\'a flood basalt province and manifested itself as major break along
the South American continental margin, similar to the Colorado and Salado
basins further south. Although sub-basalt basin structures have been reported
\citep{Eyles.G.93, Exxon.85}, there is no evidence for a large scale
continental shear zone obscured by the Paran\'a large igneous province (LIP).

In our model we have subdivided the present-day South American continent in
the following tectonic blocks, partly following previous authors
\citep[Fig.~\ref{fig:rigidplates}; e.g.][]{Moulin.ESR.09, Torsvik.GJI.09,
Macdonald.MPG.03, Nuernberg.Tectp.91, Unternehr.Tectp.88}:
\begin{enumerate}
\item The main South American Platform (SAm), extending from the Guyana Craton in the
North to the Rio de la Plata Craton in the South, with the exception of the NE Brazilian
Borborema Province.
\item The NE Brazilian Borborema Province block (BPB).
\item The Salado Subplate, located between the Salado and Colorado Basins.
\item The \emph{Patagonian extensional domain} south of the Colorado Basin, composed of the 
rigid Pampean Terrane, North Patagonian Massif, Rawson block, San Julian block,
the Deseado block, and Malvinas/Falkland block, and the Maurice Ewing Bank
extended continental crust. 
\end{enumerate}

\subsubsection{Northeast Brazil}
\label{ssub:northeast_brazil}

The Borborema Province block (BPB; Figs.~\ref{fig:rigidplates} and
\ref{fig:sam}) is located in NE Brazil and separated from the South American
plate along a N-S trending zone extending from the Potiguar Basin on the
eastern Brazilian Equatorial Atlantic margin southwards to the
Rec\^oncavo-Tucano-Jatob\'a rift (RTJ, Fig.~\ref{fig:sam}). It is an
exception in the otherwise tectonically stable South American plate and a set
of isolated Early Cretaceous rift basins (e.g.~Araripe and Rio do Peixe
basins) as well as abundant evidence of reactivated Proterozoic aged basement
structures and shear zones indicates distributed, but highly localised lithospheric deformation
during the opening of the South Atlantic rift \citep{Oliveira.MPG.03,
Matos.AGU-GM.00, Mohriak.AGU-GM.00, Chang.Tectp.92, Matos.T.92a,
Milani.Tectp.88, Castro.Tect.87}. Rifting commenced in the Latest
Jurassic/Berriasian and lasted until the Mid-Barr\^emian, well documented
through extensive hydrocarbon exploration \citep{Matos.AGU-GM.00,
DarrosDeMatos.GSLSP.99, Szatmari.G.99, Magnavita.Tect.94, Chang.Tectp.92,
Milani.Tectp.88}.

Similar to the smaller tectonic blocks in the Benoue Trough region, we regard
the lithospheric deformation affecting this block caused by the motions of
the larger surrounding tectonic plates. The Borborema province is
``crushed'' during the early translation of South America relative to
Africa with extension in the West Congo cratonic lithosphere localised along
existing and reactivated basement structures leading to small, spatially confined
basins such as the Araripe, Rio do Peixe, Iguatu and Lima Campos.

We model the rifting in the Rec\^oncavo-Tucano-Jatob\'a and Potiguar basins
by allowing for $\approx$40 and 30\,km extension, respectively, through
relative motions between South America and the Borborema Province block
between 143\,Ma and 124\,Ma (Mid-Barr\^emian).

\subsubsection{Southern South America}
\label{ssub:southern_south_america}

The WNW-ESE striking Punta del Este Basin, the genetically related Salado
Basin adjacent to the South and the ENE-WSW trending Santa Lucia
Basin/Canelones Graben system, delimit our South American Block towards the
South \citep[Fig.~\ref{fig:sam};][]{Soto.MPG.11, Jacques.JGSL.03b,
Kirstein.JPet.00, Stoakes.AAPG-B.91, Zambrano.JGR.70}. Well data supports the
onset of syn-rift subsidence around the Latest Jurassic/Early Cretaceous and
post-rift commencing at Base Aptian \citep{Stoakes.AAPG-B.91}. The
rift-related structural trend is predominantly parallel to the basin axis,
indicating a NNE-SSW directed extension. Sediment thicknesses reach 6\,km
with crustal thicknesses around 20--23\,km \citep{Croveto.BIFG.07} yielding
stretching factors of around 1.4. We have implemented 40\,km of NE-SW
transtension between 145\,Ma to Base Aptian for the eastern part of the
basin, which is assumed to have been linked towards the west by a zone of
diffuse deformation with the General Levalle basin. We have split the
southern Rio de la Plata craton along the syntaxis of the Salado/Punta del
Este Basin between the South American block and the Salado Sub-plate.

The rigid Salado block contains the Precambrian core of the Tandilia region
and the Paleozoic Ventania foldbelt \citep{Pangaro.MPG.12, Ramos.JSAES.08}
and is delimited by the Late Jurassic/Early Cretaceous-aged Colorado,
Macach\'in, Laboulaye/General Levalle and San Luis basins in the South,
Southwest and West, respectively \citep[Fig~\ref{fig:sam};][]{Pangaro.MPG.12,
Franke.GJI.06, Webster.AAPG-B.04, Urien.AAPG-M62.95, Zambrano.JGR.70}.
Basement trends deduced from seismic and potential field data indicate,
similar to the Salado Basin, E-W trending rift structures, orthogonal to the
SARS and point to to N-S directed extension/transtension \citep[][J. Autin
personal communication, 2012]{Pangaro.MPG.12, Franke.GJI.06}.
\citet{Pangaro.MPG.12} estimate around 45\,km (20\%) N-S directed extension
for the Colorado basin. Our model assumes $\approx$50\,km of NE-SW directed
transtension for the basin from 150\,Ma to Base Aptian when relative motions
between Patagonian Plates and South America cease \citep{Somoza.EPSL.08}.

The Colorado Basin marks the transition between the blocks related to the
South American Platform and the Patagonian part of South America
\citep{Pangaro.MPG.12}, which we here summarise as Patagonian extensional
domain. The Patagonian lithosphere south of the Colorado Basin is composed of
a series of amalgamated magmatic arcs and terranes with interspersed
Mesozoic sedimentary basins \citep{Ramos.JSAES.08,Macdonald.MPG.03,
Ramos.E.88,Forsythe.JGS.82}. For the purpose of this paper, the North
Patagonian Massif, Rawson Block, Deseado Block and Malvinas/Falkland Island
Block are not separately discussed as deformation in the Colorado and Salado
basins largely accounts for clockwise rotation of the Patagonian South
America during the Late Jurassic to Aptian.

\subsubsection{The Gastre Shear Zone}
\label{ssub:the_gastre_shear_zone}

The Gastre shear zone is used in previous plate tectonic models as major
intracontinental shear zone, separating Patagonian blocks from the main
South American plate \citep{Torsvik.GJI.09,Macdonald.MPG.03}. However, no
substantial transtensional or transpressional features along this proposed
faults zone are recognised in this part of Patagonia, nor is the geodynamic
framework of southern South America favouring the proposed kinematics. A
detailed geological study of the Gastre Fault zone area lead
\citet{vonGosen.JSAES.04} to conclude that there is no evidence for a late
Jurassic--Early Cretaceous shear zone in the Gastre area. Our model does not
utilise a Gastre Shear Zone to accommodate motions between the South American
and Patagonian blocks.

\section{Plate reconstructions}
\label{sec:reconstructions}

The plate kinematic evolution of the South Atlantic rift and associated
intracontinental rifts preserved in the African and South American plates, is
presented as self-consistent kinematic model with a set of finite rotation
poles (Table~\ref{tab:rotationparams}). In the subsequent description of key
timeslices we refer to Southern Africa (SAf) fixed in present day position.
We will focus on the evolution of the conjugate South Atlantic margins. For
paleo-tectonic maps in 1\,Myr time intervals please refer to the electronic
supplements.

\subsection{Kinematic scenarios}
\label{sub:evaluating_different_kinematic_scenarios}

\begin{figure*}[t]
	\vspace*{2mm}
	\begin{center}
		\includegraphics[width=12cm]{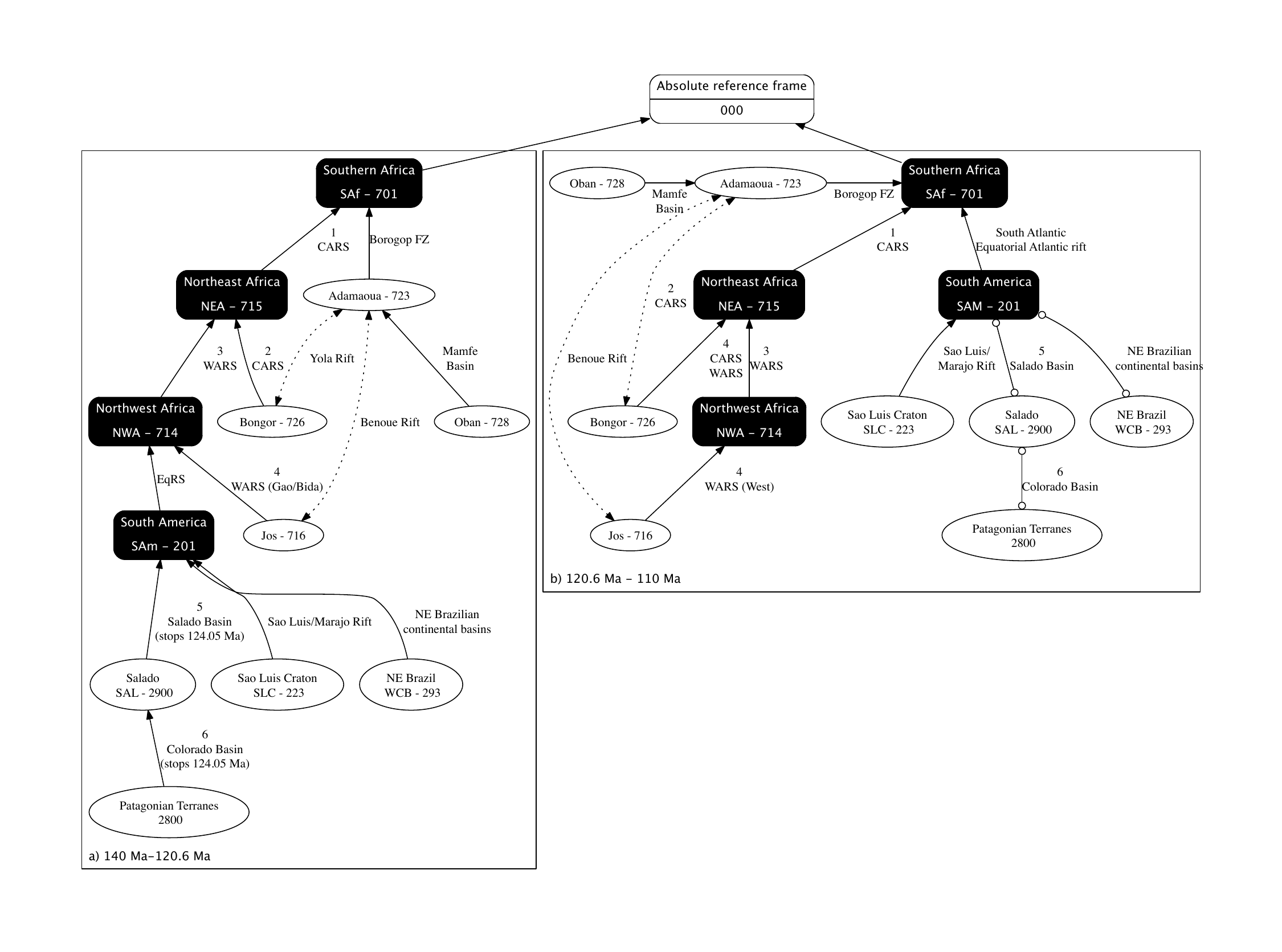}
	\end{center}
	\caption{Plate circuits used for the plate model for two main stages: a) 140--120.6\,Ma, and b) 120.6--110\,Ma for the main lithospheric plates. Integers indicate the Plate-ID number. Integers on connection lines refer deforming zones (Tab.~\ref{tab:extnestimates}). Dashed lines indicate areas where extension is indirectly constrained by plate circuit, connections with dots indicate no relative motion. Plate abbreviations }
	\label{fig:platecircuit}
\end{figure*}

Plate motions are expressed in the form of plate circuits or
rotation trees in which relative rotations compound in a time-dependent,
non-commutative way. The core of our plate tectonic model is the quantified
intraplate deformation which allows us to indirectly model the time-dependent
velocities and extension direction in the evolving South Atlantic rift. Between initiation
and onset of seafloor spreading, the plate circuit for the South America
plate is expressed by relative motions between the African sub-plates
(Fig.~\ref{fig:platecircuit}). The kinematics of rifting are
well constrained through structural elements and sedimentation
patterns, however, the timing of extension carries significant
uncertainties due to predominantely continental and lacustrine sediment
infill. Limited direct information from drilling into the deepest
parts of these rifts is publicly available. The regional evolution allows for a
relatively robust dating of the onset of deformation at the Base Cretaceous 
in all major rift basins \citep[e.g.][]{Janssen.GSA-B.95}, whereas the onset of post-rift subsidence in 
the CARS and WARS, is not as well constrained and further complicated through
which have experienced subsequent phases of significant reactivation 
\citep[e.g.][]{Guiraud.JAfES.05}. 

The design of our plate circuit (Fig.~\ref{fig:platecircuit})
implies that the timing of rift and post-rift phases significantly affect the resulting
relative plate motions between South America and Southern Africa.
We have tested five alternative kinematic scenarios by varying the duration 
of the syn-rift phase in the African intracontinental rifts, and 
along the Equatorial Atlantic margins to evaluate the temporal
sensitivity of our plate model (Tab.\ref{tab:kinmodels}). 
The onset of syn-rift was changed between 140\,Ma and 135\,Ma in the CARS 
and WARS, along with the end of the syn rift phase ranging between 110\,Ma 
and 100\,Ma. For the Equatorial Atlantic rift, we evaluated syn-rift phase 
onset times between 140\,Ma and 132\,Ma (Top Valanginian). Additionally, 
we included the plate model of \citet{Nuernberg.Tectp.91} 
with forced breakup at 112\,Ma \citep[``NT91'';][]{Torsvik.GJI.09} in our 
comparison. Flowlines for each alternative scenario were 
plotted and evaluated against observed lineations and fracture zone patterns 
of filtered free-air gravity. The implied extension history for the 
conjugate South Atlantic passive margins was used as a primary criterion 
to eliminate possible alternative scenarios. In the preferred model PM1,
rifting along SARS, CARS, WARS, and EqRS starts simultaneously 
around the early Berriasian (here: 140\,Ma). It satisfies geological
(timing and sequence of events) and geophysical observations
(alignment of flowlines with lineaments identified in the gravity data) for
all marginal basins. Alternatively tested models introduced kinematics
such as transpression in certain parts of the margins for which are not supported 
by the geological record. All tested scenarios (PM1-PM5), however, 
show a good general agreement, confirming the robustness of our methodology of constructing
indirect plate motion paths for the evolution of the SARS through accounting
for intraplate deformation.

The largest difference between our model preferred model PM1 and model NT91 are the start of rifting and the implied 
kinematic history for the initial phase of extension. 
Relative motions between SAm and the African plates commences at 131\,Ma in NT91. 
This results in higher extensional velocites and strain rates in NT91 for all 
extensional domains due to a shorter duration ($\Delta t$~=~12\,Myrs) between the
onset of plate motions and key tiepoint at Chron M0 (the same across
all models). 

% Rifting between
% the NE Brazilian Borborema Province Block and the Douala/Benoue Microplates
% occurred much later compared to the main SARS. Douala and Kribi-Campo Basins (conjugate to the Brazilian
% Pernambuco-Paraiba margin; Figs.~\ref{fig:afr} and \ref{fig:sam}) Precambrian basement rocks have been drilled,
% overlain by Barr\^emian to Aptian in age, approximately 15\,Myrs younger than
% the earliest known syn-rift deposits in the SARS
% \citep{Brownfield.USGS-B.06b, Turner.Tectp.03}.
% as most of the
% pre-Barr\^emian extensional deformation was taken up by the
% Rec\^oncavo-Tucano-Jatob\'a, Araripe and Potiguar rifts. 
% In the , the
% stratigraphy indicates a slightly later onset of subsidence an rifting
% compared to the other margin segments (Gabon/Sergipe
% Alagoas to Campos/Kwanza). 

\begin{figure}[t]
\vspace*{2mm}
\begin{center}
	\includegraphics[width=8.3cm]{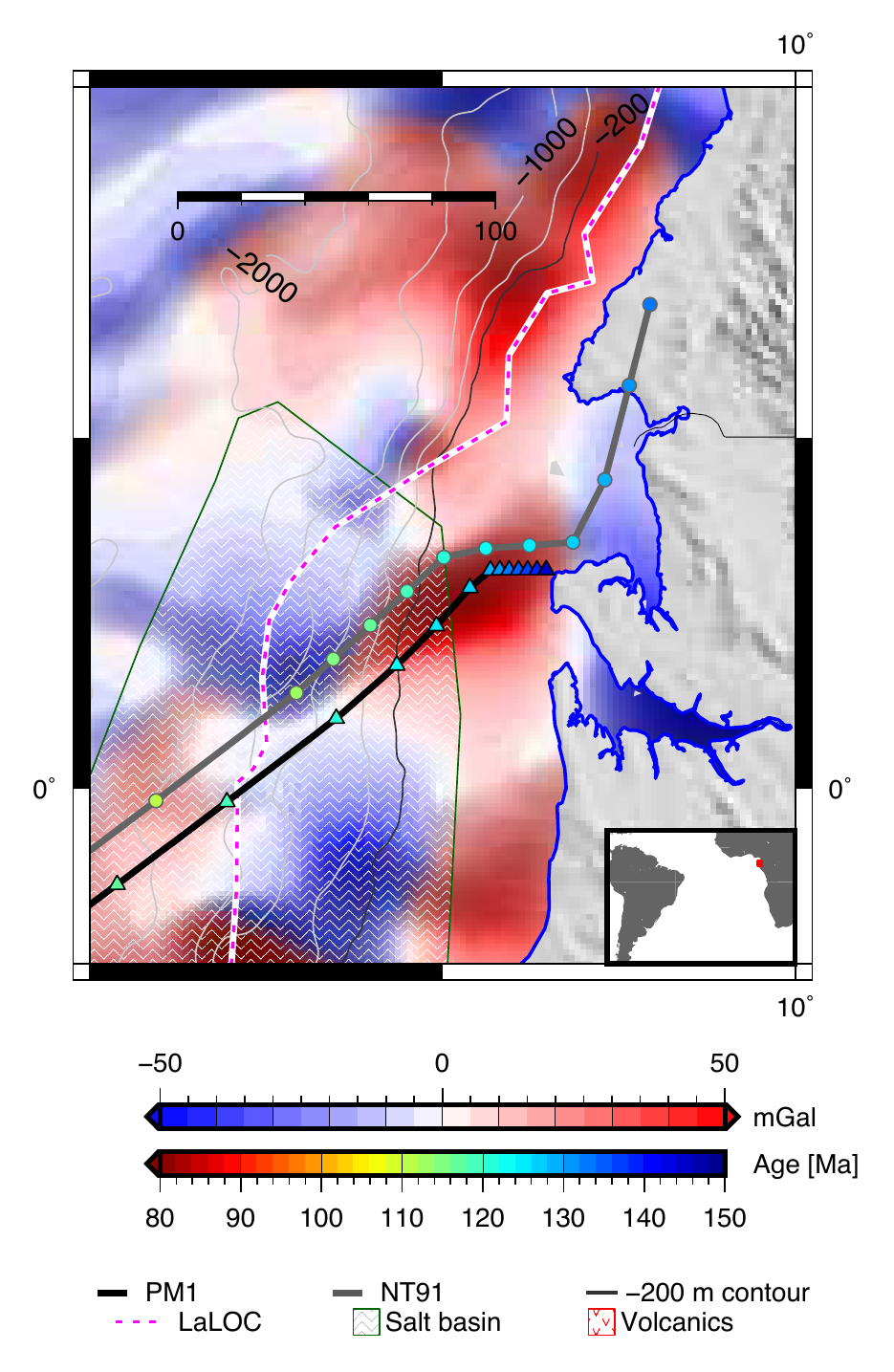}	
	\end{center}
		\caption{
		Age-coded flowlines plotted on filtered free air gravity basemap (offshore) for the North Gabon/Rio Muni segment of the African margin. Initial phase flowline direction of our preferred model (PM1, thin black line with circles) is perpendicular to the trend of the present-day coastline, striking NW-SE and only turns parallel to oceanic transform zones (trending SW-NE) after $\approx$ 126 ma. Note that models PM4 \& PM 5 induce a significant amount of compression for the intial phase of relative plate motions and do not agree with geological observations from the area. Relative motions between SAf and SAm in Model NT91 commence at 131 ma, Circles are plotted in 2 Myr time intervals starting at 144 ma and from 132 for NT91 \citep{Nuernberg.Tectp.91, Torsvik.GJI.09}. Legend abbreviations: LaLOC - Landward limit of oceanic crust, C. - Contour. Free air gravity anomalies \citep{Sandwell.JGR.09} filtered with 5th order Butterworth lowpass filter with 35 km wavelength.
		\label{fig:flowgabon}
		}  
\end{figure}

In the northern Gabon region, the flowlines for
model NT91 indicate a rapid initial NNE-SSW translation of South America relative to Africa
by about 100\,km for the time from 131--126\,Ma (Fig.~\ref{fig:flowgabon}). The resulting
transpression along the northern Gabon/Rio Muni margin during this time interval, is not evidenced 
from the geological record 
\citep[e.g.][]{Brownfield.USGS-B.06b, Turner.Tectp.03}. This initial extension
phase is followed by a sudden E-W kink in plate motions from 126\,Ma to
118\,Ma before the flowlines turn SW-NE, parallel to our model(s) for the
time of the CNPS. Our modeled extension history for the  Gabon margin implies
an initial 40-60 km E--W directed rifting between South America and Africa until 127 Ma, 
followed by a 40$^{\circ}$~rotation of the extension direction and subsequent increase
in plate velocities.

\begin{figure}[t]
\vspace*{2mm}
\begin{center}
	\includegraphics[width=8.3cm]{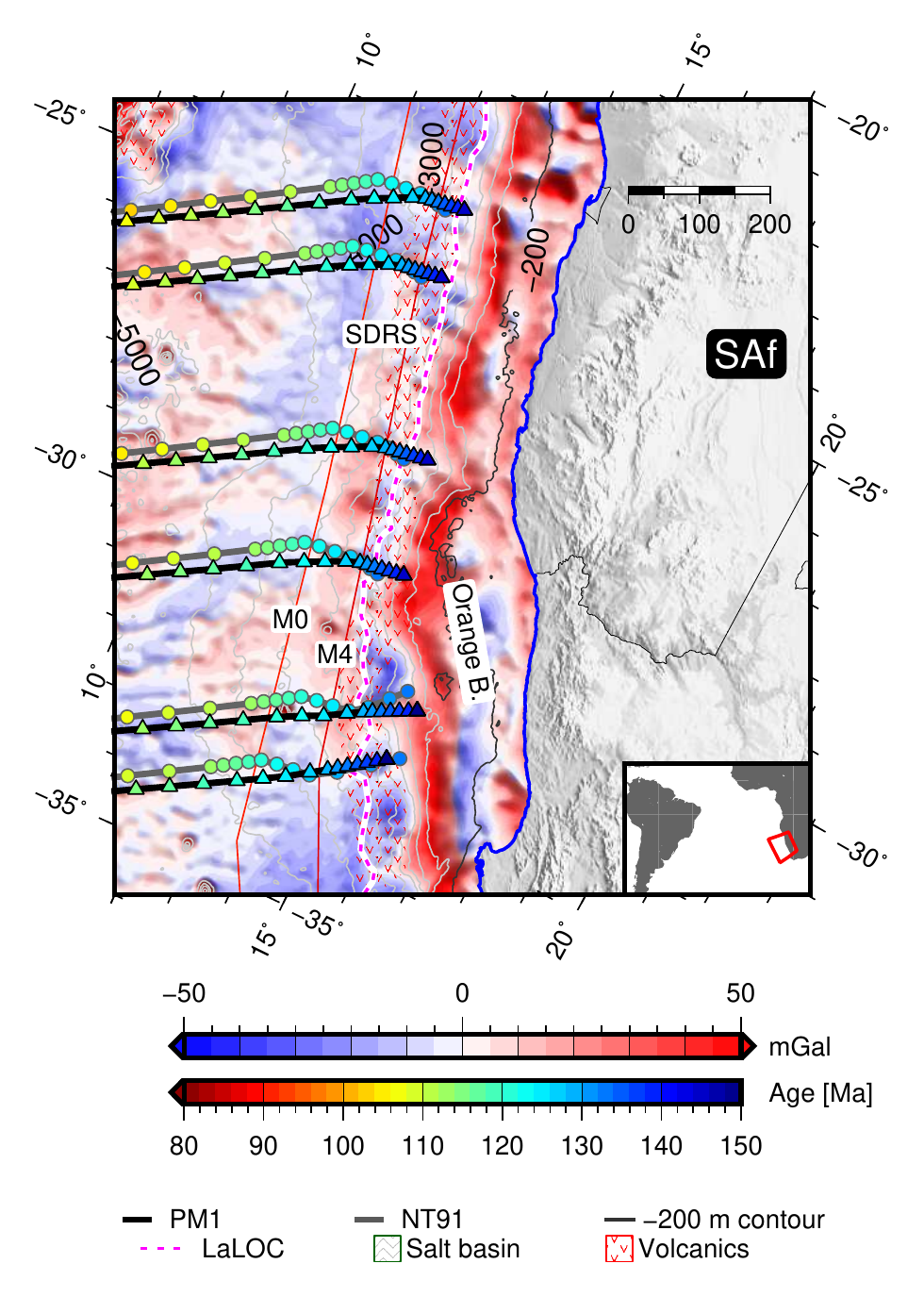}	
	\end{center}
		\caption{
		% TODO IMAGE NEEDS MORE LABELS
		Age-coded flowlines plotted on filtered free air gravity basemap (offshore) for the Orange Basin segment along the West African margin. Early phase opening is oblique to present day margin, with oblique initial extension of SAm relative to SAf (4 northern flowlines) in WNW-ESE direction  and intial extension between Patagonian South American blocks and SAf in SW-NE direction (southern 2 flowlines). Note that the Orange Basin is located between the two divergent flowline populations and that  a positive gravity anomaly is contemporaneous with  the inflection point at 126.57 ma (M4) and associated velocity increase and extension direction change (cf. Fig.~\ref{fig:velplot}) along the margin. Legend/symbology and filter as in Fig.~\ref{fig:flowgabon}.
		\label{fig:flownamib}
		}  
\end{figure}

Along the Nambian margin, predicted initial extension directions for the 
relative motions between South America--Africa and Patagonian terranes--Africa 
diverge in the central Orange Basin, conjugate to the Colorado
and Salado Basins (Fig.~\ref{fig:flownamib}). Northward from here, 
relative motions for South America--Africa are directed WNW--ESE, 
while the Patagonian Terranes--Africa motions south of the central 
Orange Basin imply opening of the southernmost SARS in ENE-WSW direction. 
This is a result of the relative clockwise rotation of the Patagonian Terranes 
away from South America ue to the rifting in the Salado/Punta del Este and Colorado basins
and accordance with structural observations from the southern Orange Basin
(H.~Koopmann, personal communication, 2012). These model predictions are supported 
by the trend of the main gravity lineaments in contrast to the plate motions paths of 
model NT91 which are not reconcilable with gravity signatures (Fig.~\ref{fig:flownamib}). 

\begin{figure}[t]
\vspace*{2mm}
\begin{center}
	\includegraphics[width=8.3cm]{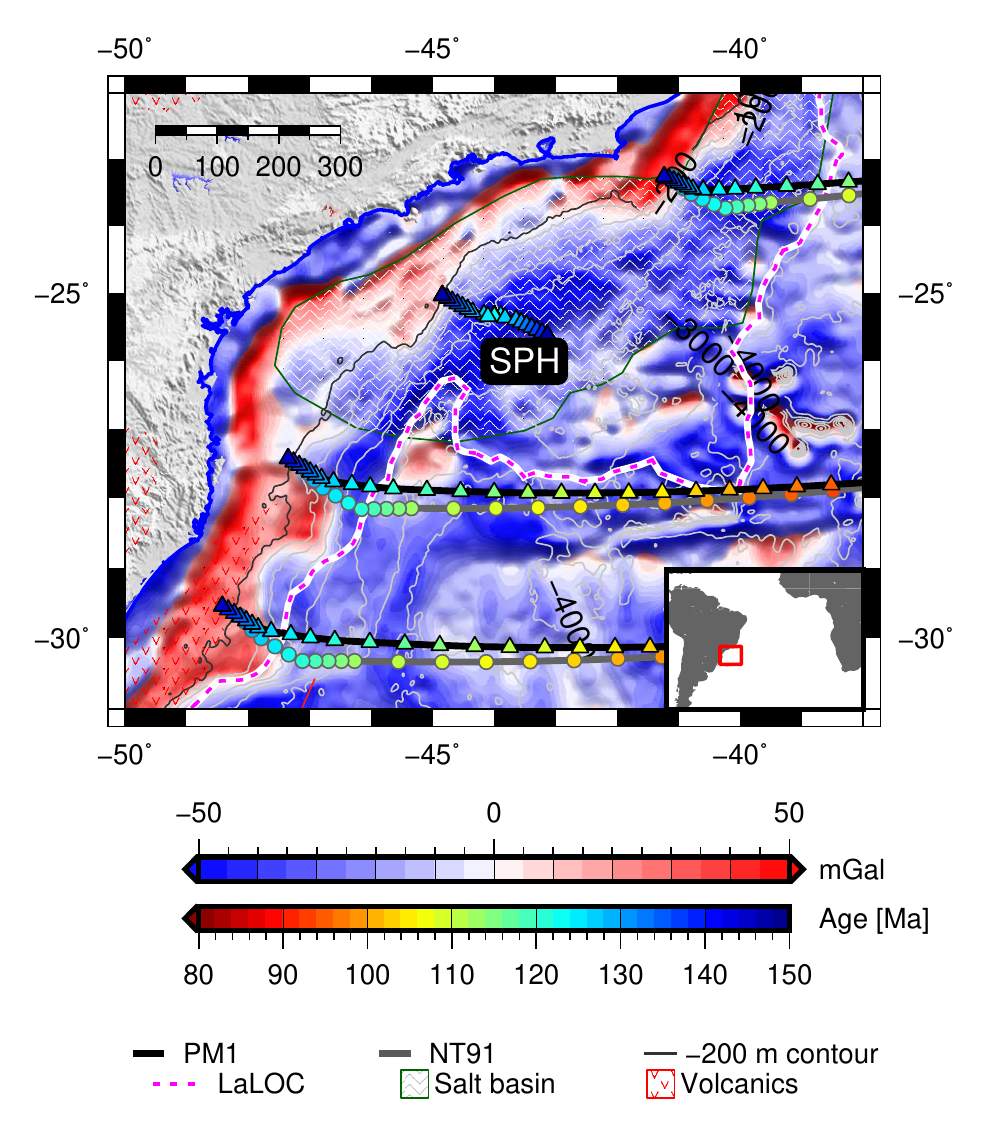}	
	\end{center}
		\caption{
		Age-coded flowlines plotted on filtered free air gravity basemap (offshore) for the Santos Basin segment on South American margin. SPH- S\~ao Paulo High in the outer Santos basin. Note that initial NW-SE relative extensional direction as predicted by the plate model is perpendicular to the Santos margin hingeline \citep[cf.][for details on the crustal structure in along the inner Santos basin margin]{Meisling.AAPG-B.01} and does conform to observed structural patterns in the extended continental crust of the SPH \citep[cf.][]{Chang.URL.12}. Legend/symbology and filter as in Fig.~\ref{fig:flowgabon}
		\label{fig:flowsantos}
		}  
\end{figure}

In the Santos basin our preferred model PM1 predicts an initial
NW-SE directed extension in the proximal part, oriented
orthogonally to the main gravity gradients, Moho topography and proximal
structural elements \citep[Fig.~\ref{fig:flowsantos}; e.g.][]{Stanton.Tectp.10, Chang.URL.12,
Meisling.AAPG-B.01}. In the western Santos basin we model the initial
extension phase to be focussed between the S\~ao Paulo High/Africa and South America,
until the onset of the third extensional phase around the Barr\^emian/Aptian
boundary, a time when the inferred oceanic Abimael spreading becomes extinct
\citep[Figs.~\ref{fig:platecircuit} and
\ref{fig:flowsantos};][]{Scotchman.GSL-PGC.10} and the S\~ao Paulo High is
translated from the African to the South American plate. 

\subsection{Fit reconstruction and the influence of Antarctic plate motions}

\label{sub:fit_reconstruction}

\begin{figure*}[t]
\vspace*{2mm}
\begin{center}
\includegraphics[width=11cm]{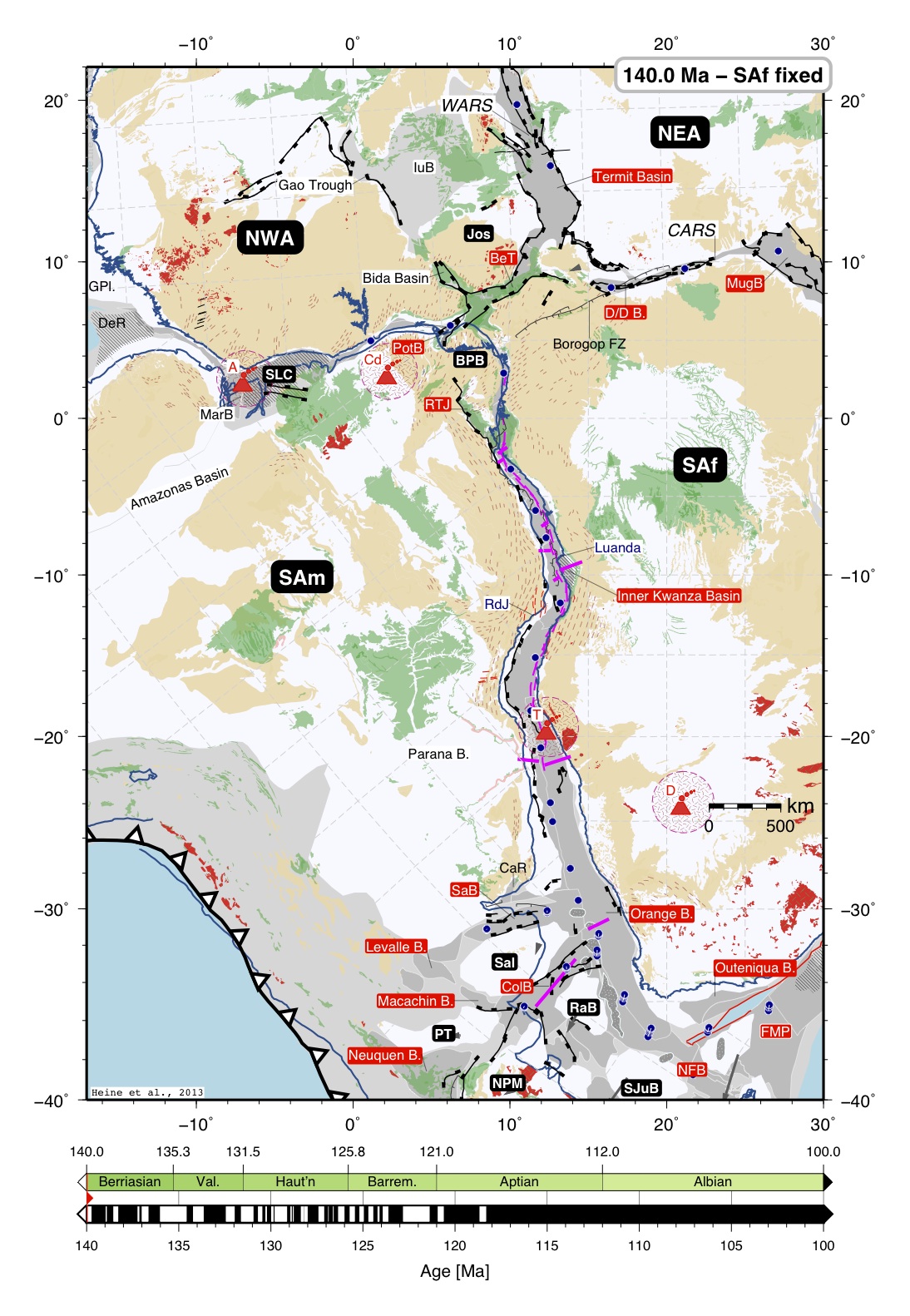}	
 	\end{center}
 		\caption{
 		Plate tectonic reconstruction at 140\,Ma, Africa fixed in present-day coordinates. For map legend see Fig.~\ref{fig:maplegend}, bottom color scales indicate geological stages and magnetic reversals with red diamond indicating time of reconstruction, abbreviations here: T - Tithonian, Val. - Valaginian, Haut'n - Hauterivian, Barrem. -  Barr\^emian. Restored continental margin is indicated by dashed magenta colored line, restored profiles (Fig.~\ref{fig:crosssections}) by solid magenta lines. Rigid lithospheric blocks denoted by black labels: SAf - Austral African Plate, BoP - NE Brazilian Borborema Province plate, Jos - Jos Plateau Subplate, NEA - NE African Plate, NPM - North Patagonian massif, PT - Pampean Terrane, RaB - Rawson Block, Sal - Salado Subplate, SAm - main South American Plate, SJuB - San Julian Block, SLC - S\~ao Luis Craton Block. Actively extending basins are indicated by red background in label, postrift basins are indicated by light gray background in label, abbreviations: BeT - Benoue Trough, CaR - Canelones Rift, D/D B. - Doba and Doseo Basins, ColB - Colorado Basin, IuB - Iullemmeden Basin, marB - maraj\'o Basin, MugB - Muglad Basin, NFB - North Falkland Basin, PotB - Potiguar Basin, TRJ - Reconavo, Jatoba, Tucano Basins, SaB - Salado Basin. Other abbreviations: RdJ - Rio de Janeiro, B. - Basin, GPl. - Guyana Plateau, DeR - Demerara Rise. Present-day hotspots are shown as volcano symbol and plotted with 400 km diameter (dashed magenta colored circle with hachured fill), assuming that they are stationary over time: A - Ascension, B - Bouvet, Ch - Chad/Tibesti, Ca - Mount Cameroon, Cd - Cardno Seamount, D - Discovery, T - Tristan da Cu\~nha.\label{fig:140}
		}  
\end{figure*}

\begin{figure*}[t]
\vspace*{2mm}
\begin{center}
	\includegraphics[width=12cm]{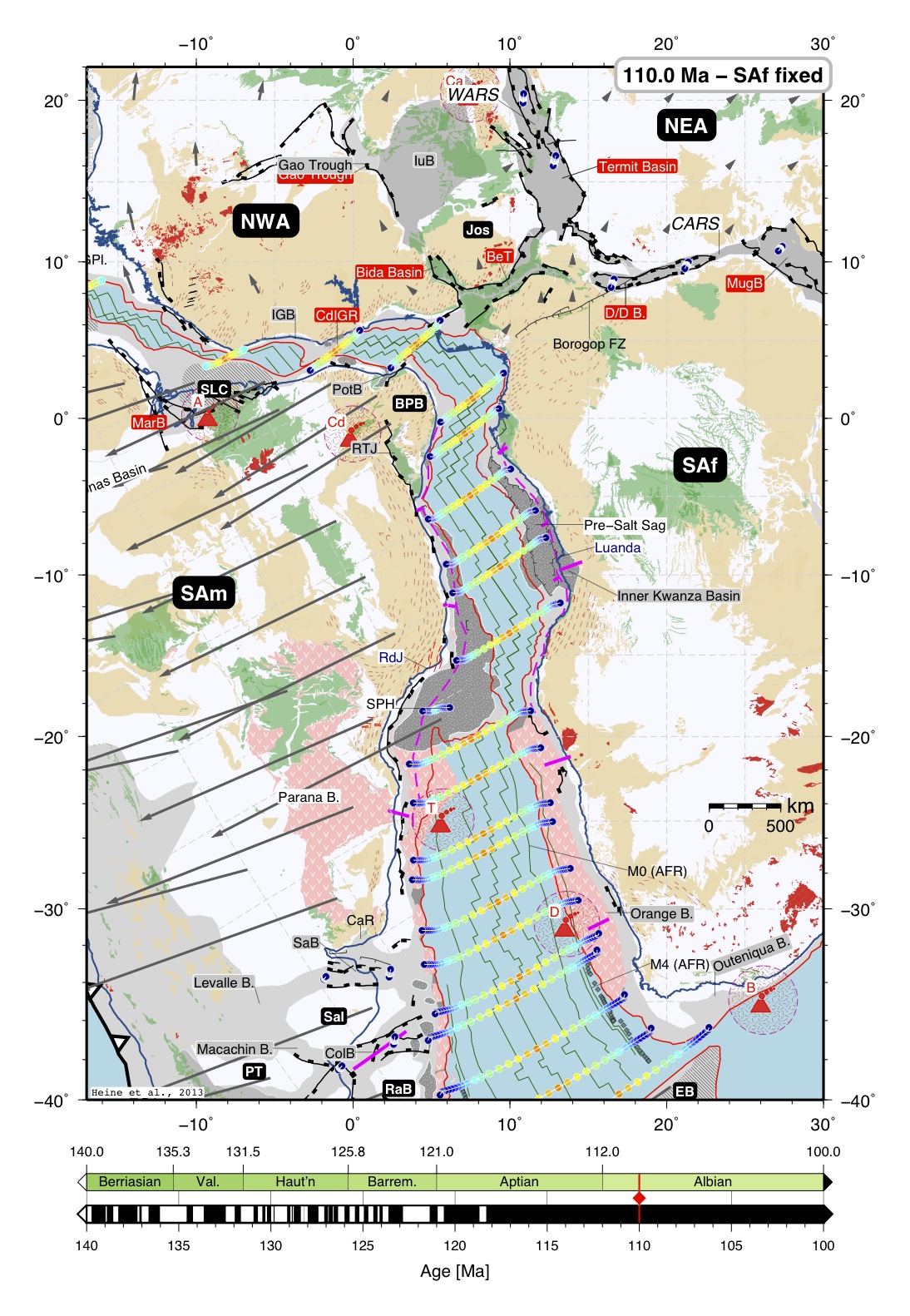}	
	\end{center}
		\caption[Reconstruction at 110\,Ma]{
		Plate tectonic reconstruction at 110\,Ma, with Africa fixed in present-day coordinates. For map legend see Fig.~\ref{fig:maplegend}, abbreviations as in Fig.~\ref{fig:140}. 
		Full continental separation is achieved at this time, with narrow oceanic gateways now opening between the C\^ote d'Ivoire/Ghana Ridge and the Piaui-C\'eara margin in the proto-Equatorial Atlantic and between the Ewing Bank and Aghulas Arch in the southernmost South Atlantic. Deformation related to the break up between Africa and South America in the African intracontinental rifts ceases in post-Aptian times.
		Towards the Top Aptian, break up between South America and Africa has largely been finalised. The only remaining connections are between major offset transfer faults in the Equatorial Atlantic rift and between the outermost Santos Basin and the Benguela margin where a successively deeping oceanic gateway between the northern and southern Proto-South Atlantic is proposed. Seafloor spreading is predicted for the conjugate passive margin segments such as the Deep Ghanaian Basin (DGB). The stage pole rotation between SAm and NWA predicts compression alon gthe CdIGR in accordance with observed uplift during this time \citep[e.g.][]{Pletsch.JSAES.01, Clift.JGSL.97}. Abbreviations: CdIGR - C\^ote d'Ivoire-Ghana Ridge, EB - Maurice Ewing Bank.
		\label{fig:110}
		}  
\end{figure*}

The fit reconstruction (Fig.~\ref{fig:140}) is generated by restoring the
pre-rift stage along the intraplate WARS and CARS for the African sub-plates,
and in the Rec\^oncavo-Tucano-Jatob\'a, Colorado and Salado Basins for the
South American sub-plates using estimates of continental extension (see
Sect.~\ref{sec:tectonic_elements_rigid_blocks_and_deforming_zones}). We then
use area balanced crustal-scale cross sections along the South Atlantic
continental margins, (Sect.~\ref{sub:passive_margins},
Fig.~\ref{fig:crosssections}) in combination with the restored margin
geometry published by \citet{Chang.Tectp.92} to construct pre-rift
continental outlines. Margins along the Equatorial Atlantic are generally
narrow \citep{Azevedo.PhD.91} and associated with complex transform fault
tectonics and few available published crustal scale seismic data, the
location of the LaLOC here is only based on potential field data, with the 
expection of the Demerara Rise-French Guiana-Foz do Amazon segment 
where we have utilised \citet{Greenroyd.GJI.08, Greenroyd.GJI.07}. We 
allow on average 100\,km of overlap between the present-day continental
margins in the EqRS (Fig.~\ref{fig:overlap}), compared to a few hundred\,km
in the central part of the SARS. Total stretching estimates for profiles
across the margins for the central and southern South Atlantic segment range
between 2.3--3.8 for the 10 profiles shown in Fig.~\ref{fig:crosssections}.
South America is subsequently visually fitted against the West African and
African Equatorial Atlantic margins using key tectonic lineaments such as
fracture zone end points and margin offsets (Fig.~\ref{fig:140}).
Transpressional deformation has affected the conjugate Demerara Rise/Guinea
Plateau submerged promontories, resulting in shortening of both conjugate
continental margins during the opening of the SARS \citep{Basile.Tectp.13,
Basile.JAfES.05, Benkhelil.Tectp.95}. Our reconstructions hence show a gap of
$\approx$50\,km between the Demerara Rise and the Guinea Plateau at the
western EqRS as we have not restored this shortening (Fig.~\ref{fig:140}).

The dispersal of Gondwana into a western and eastern part was initated with
continental rifting and breakup along the incipient Somali-Mozambique-Wedell
Sea Rift \citep[e.g.][]{Koenig.JGR.06,Norton.JGR.79}. NE-SW directed displacement of
Antarctica as part of eastern Gondwana southwards relative to South America and Africa
creates an extensional stress field which affects southernmost Africa and
the present-day South American continental promontory comprised of the Ewing
Bank and Malvinas/Falkland Island and the Proto-Weddell sea from the
Mid-Jurassic onwards \citep{Koenig.JGR.06, Macdonald.MPG.03}.
% The Patagonian Terranes (Fig.~\ref{fig:rigidplates}) are separated by rift-related
% basins 
Our pre-rift reconstruction for the Patagonian extensional domain takes into
account possible earlier phases of extension which have affected, for example,
the San Jorge, Deseado or San Jul{\'i}an Basins \citep{Jones.G3.04, Homovc.AAPG-B.01} and
hence should represent a snapshot at 140 Ma, rather than a complete fit reconstruction which
restores all cumulative extension affecting the area in Mesozoic times. 
Initial rifting of this region preceeding relative motions
between South America and Africa, is recorded by Oxfordian-aged syn-rift 
sediments in the Outeniqua Basin in South Africa
as well as subsidence and crustal stretching in the North Falkland Basin and
the Maurice Ewing Bank region \citep[Fig.~\ref{fig:sam};
][]{Jones.G3.04, Paton.BR.04, Macdonald.MPG.03, Bransden.GSLSP.99}. The overall NE-SW
extension causes a clockwise rotation of the Patagonian blocks away from SAf
and SAm commencing at $\approx$150\,Ma in our model, with an fit reconstruction 
position of the Falkland Islands just south Aghulas Arch/Outeniqua basin. 
This early motion results in the onset of crustal stretching in the North
Falkland (150 Ma), Colorado (150 Ma) and Salado (145 Ma) basins which preceeds relative motions between
the main African and South American plates (Fig.~\ref{fig:140}). The resulting
relative motion in the southern part of the SARS is NE-SW directed
transtension (Fig.~\ref{fig:flownamib}), and E-W extension between the Malvinas/Ewing Bank promontory and Patagonia. 
Our fit reconstruction for the southernmost South Atlantic places the Malvinas/
Falkland Islands immediately south of the Aghulas Arch/Outeniqua Basin. Based on 
published extension estimates for relevant basins in southern Patagonia/offshore
Argentina \citep[e.g.]{Jones.G3.04}, alternative pre-rift reconstructions placing the Falkland Islands
further east and excessive microplate rotation \citep{Mitchell.N.86} 
can not be reconciled with the currently available data. 

\subsection{Phase I: Initial opening -- Berriasian to late Hauterivian \\(140--126.57\,Ma)}

\label{sub:phase1}

Extensional deformation along the WARS, CARS and SARS is documented to start
in the Early Cretaceous (Berriasian) by the formation of intracontinental
rift basins and deposition of lacustrine sediments along the conjugate South
Atlantic margins \citep{Chaboureau.Tectp.12, Poropat.GR.12, Dupre.MPG.07,
Brownfield.USGS-B.06b, Bate.GSLSP.99, Cainelli.Epis.99, Coward.GSLSP.99,
Karner.MPG.97, Chang.Tectp.92, Guiraud.Tectp.92b}. Here, we use 140 Ma
as absolute starting age of the onset of deformation in the South Atlantic,
Central and West African rift systems (lower Berriasian in our Geek07/GTS04 hybrid timescale, Fig.~\ref{fig:timescalecomp}).
In the southern part of the SARS, SW-directed extension has already commenced in the
latest Jurassic with syn-rift subsidence in the Colorado, Salado, Orange and
North Falkland Basins \citep{Seranne.JAfES.05, Jones.G3.04, Clemson.GSLSP.99,
Maslanyj.MPG.92, Stoakes.AAPG-B.91}. Evidence for deformation and rifting
along the EqRS during the Early Cretaceous is sparse, however, indications
for magmatism and transpressional deformation are found in basins along the
margin \citep[e.g.\,Maraj\'o Basin, S\~ao Luis Rift;][]{SoaresJunior.Gcas.11,
Costa.AABC.02, Azevedo.PhD.91}. The Ferrer-Urbanos-Santos Arch is an E--W
trending basement high between the proximal parts of the Barrerinhas basin
and the onshore Parnaiba basin which was generated by transpression during
the Neocomian \citep{Azevedo.PhD.91}.

\begin{figure}[t]
\vspace*{2mm}
\begin{center}
	\includegraphics[width=8.3cm]{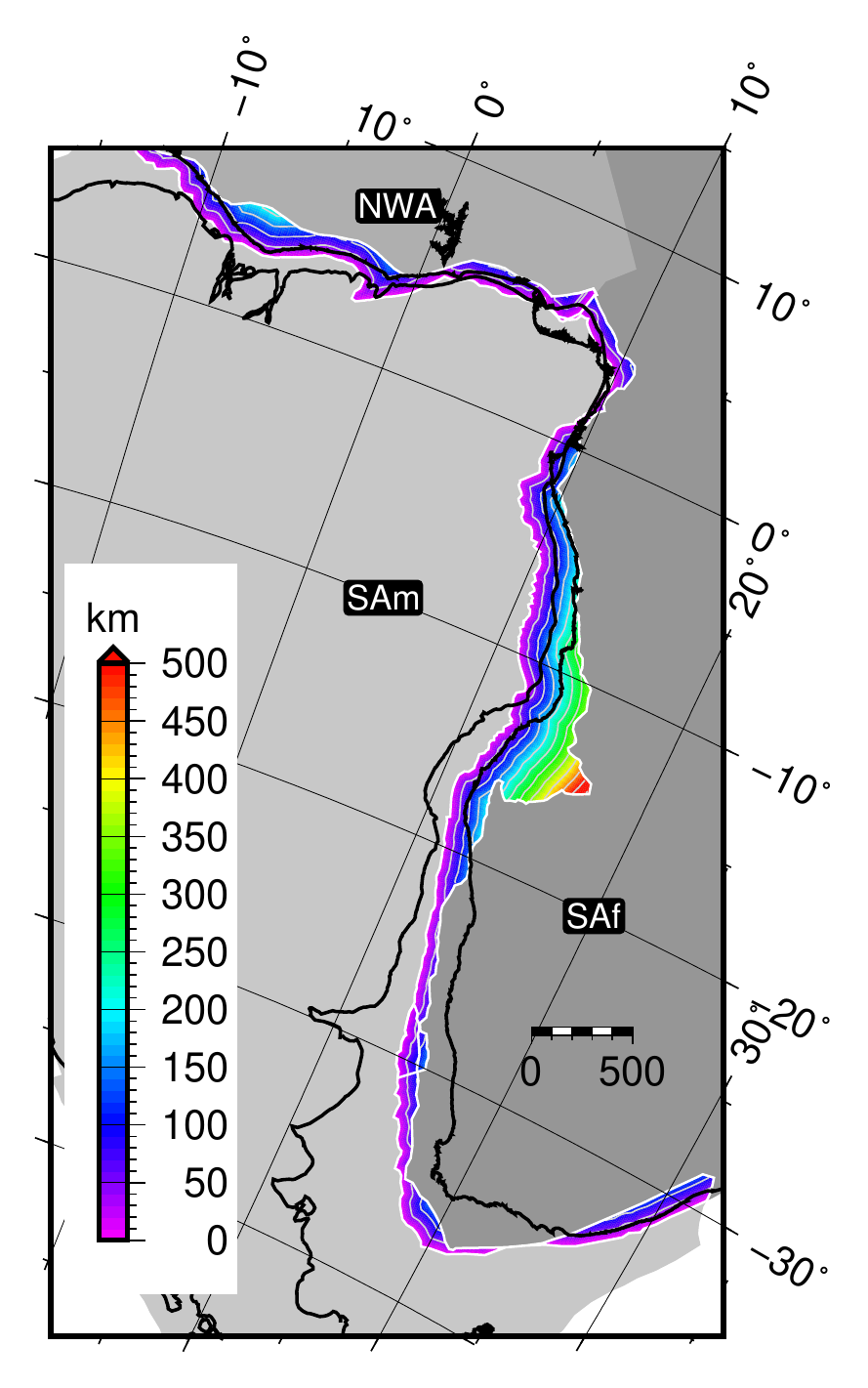}	
	\end{center}
		\caption{
		Overlap between the conjugate present-day landward limits of the oceanic crust (LaLOC) for the South America (SAm), Southern Africa (SAf) and Northwest Africa (NWA). Shown is the fit reconstruction at 143 with SAf fixed in present-day position. Spacing between contour lines is 50 km, present-day coastlines as thick black lines. Note that the position of the LaLOC along the volcanic margins in southern part of the South Atlantic rift is not well constrained. Largest overlap is created between the Campos/Santos basin margin and the southern Kwanza/Benguela margin implying up to 500 km extension for the SE'most Santos Basin.
		\label{fig:overlap}
	}
\end{figure}

\begin{figure*}[t]
\vspace*{2mm}
\begin{center}
	\includegraphics[width=12cm]{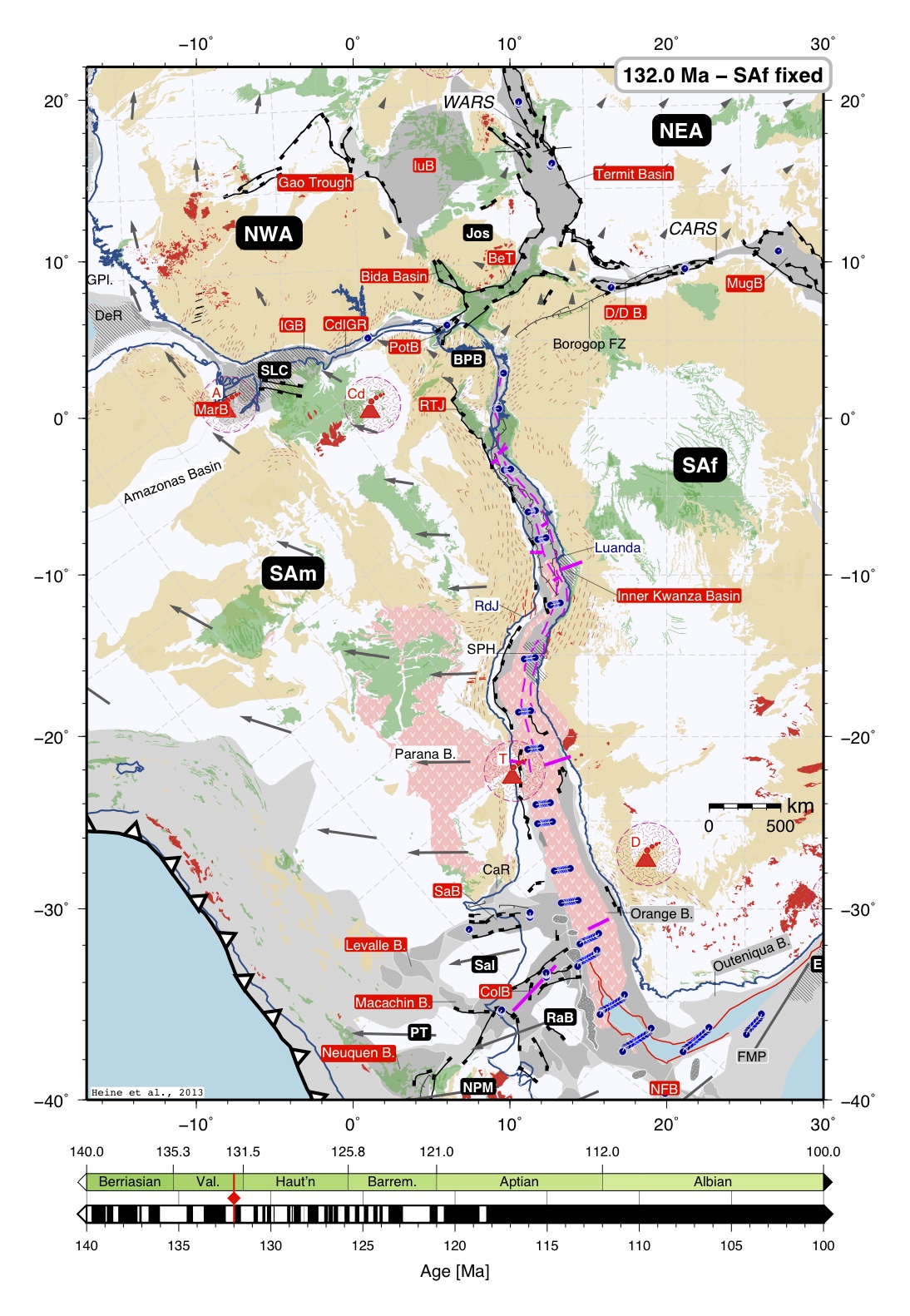}	
	\end{center}
		\caption{
		Plate tectonic reconstruction at 132\,Ma, with Africa fixed in present-day coordinates. For map legend see Fig.~\ref{fig:maplegend}, abbreviations as in Fig.~\ref{fig:140}. Note initial extension directions along the margin rotate from NW-SE in Gabon/Sergipe-Alagoas segment to W-E in Pelotas/Walvis Basin segment with increasing distance from stage pole location. Flowlines between Patagonian blocks in southern South America and southern Austral Africa indicate and initial SW-NE directed motions between these plates (cf.~Fig~\ref{fig:flownamib}). In West Africa, the Iullemmeden and Bida Basin as well as the  Gao Trough are undergoing active extension. Tectonism along the Equatorial Atlantic rift increases. Additional abbreviations: DGB - Deep Ghanian Basin, CdIGR - C\^ote d'Ivoire/Ghana Ridge and associated marginal basins.
		\label{fig:132}
		}  
\end{figure*}

Extension in all major rift basins occurs at slow rates during the initial
phase, with separation velocities between SAm and SAf around 2\,mm\,a$^{-1}$
in the Potiguar/Rio Muni segment and up to c. 20\,mm\,a$^{-1}$ in the
southernmost SARS segment, closer to the stage pole equator
(Fig.~\ref{fig:velplot}). Extensional velocities in the magmatically dominated
margin segments of the southern South Atlantic are well above 10\,mm\,a$^{-1}$.
The predominant extension direction changes from
NW-SE in the northern SARS segment to WNW-ESE in the southern part, and
SSW-NNE for the conjugate Patagonia-SAf segment (Figs.~\ref{fig:velplot} and
\ref{fig:140}--\ref{fig:132}. Along the central SARS segment, this phase
corresponds to \citet{Karner.MPG.97}'s ``Rift Phase 1'' with broadly
distributed rifting.

Flowlines derived from the plate model for the early extension phase
correlate well with patterns observed in the free air gravity field along the
proximal parts of the margins, such as in the northern Orange, and inner
Santos Basins (Figs.~\ref{fig:flownamib}, \ref{fig:flowsantos}). Along NE--SW
trending margin segments, such as Rio Muni/Sergipe Alagoas and
Santos/Benguela, the extension is orthogonal to the margin, whereas oblique
rifting occurs in other segments. In this context we note that the strike of
the Taubat\'e Basin in SE Brazil (``Tau'' in Fig.~\ref{fig:sam}) as well as
observed onshore and offshore extensional structures and proximal Moho uplift
in the Santos Basin \citep{Stanton.Tectp.10, Fetter.MPG.09a,
Meisling.AAPG-B.01, Chang.Tectp.92} are oriented orthogonal to our modeled
initial opening direction of the South Atlantic rift. The
Rec\^oncavo-Tucano-Jatob\'a (RTJ) rift opens as we model the BPB attached to
the West African Congo Craton and Adamaoua subplate, with SAm being relatively
displaced northwestward. Here, the Pernambuco shear zone acts as continuation
of the Borogop Fault Zone into NE Brazil \citep{DarrosDeMatos.GSLSP.99}.
Potiguar Basin and Benoue Trough make up one depositional axis while the RTJ
and inner/northern Gabon belong to the central South Atlantic segment.

Around 138\,Ma (Mid-Berriasian), break up and seafloor spreading starts in
isolated compartments between the Rawson Block and the continental margin
south of the Orange Basin. The Tristan da Cu\~nha hotspot has been located beneath
the Pelotas-Walvis segment of the SARS since 145\,Ma
(Figs.~\ref{fig:140}--\ref{fig:132}, and paleo-tectonic maps in electronic
supplement) and likely caused significant alteration of the lithosphere
during the early phase rifting, with the eruption of the Paran\'a-Etendeka
Continental Flood basalt province occurring between 138--129\,Ma
\citep{Peate.AGU-GM.97, Stewart.EPSL.96, Turner.EPSL.94}. While we have not
included a temporary plate boundary in the Paran\'a Basin in our model,
evidence from the Punta Grossa and Paraguay dyke swarms
\citep{Peate.AGU-GM.97, Oliveira.RBG.89} points to NE-SW directed extension,
similar to the ``Colorado Basin-style'' lithospheric extension orthogonal to
the main SARS \citep{Turner.EPSL.94}.
Emplacement of massive seaward dipping
reflector sequences dominate the early evolution of the southern SARS segment
and are well documented by seismic and potential field data \citep[e.g.]{Blaich.JGR.11, 
Moulin.ESR.09, Bauer.JGR.00, Gladczenko.JGSL.97}. While the delineation 
of the extent of continental crust along volcanic margins remains contentious,
we model a near-synchronous transition to seafloor spreading south of the Walvis Ridge/Florianopolis High
around 127 Ma (Figs.~\ref{fig:132}--\ref{fig:125} and supplementary material).

Between 136--132\,Ma, the plume center is located
beneath the present-day South American coastline in the northern Pelotas
Basin. Our reconstructions indicate that the westernmost position of a
Tristan plume (assumed to be fixed) with a diameter of 400\,km is more than
500\,km east of the oldest Paran\'a flood basalts, at the northern and
western extremities of the outcrop \citep{Turner.EPSL.94}. Oceanic magnetic
anomalies off the Pelotas/northern Namibe margins indicate asymmetric
spreading with a predomiant accretion of oceanic crust along the African side
\citep{Moulin.ESR.09, Rabinowitz.JGR.79}, which in our reconstructions can be
explained through plume-ridge interactions of high magma volume fluxes and
relatively low spreading rates
\citep[Fig.~\ref{fig:velplot};][]{Mittelstaedt.EPSL.08}. The S\~ao Paulo --
R\'io de Janeiro coastal dyke swarms, emplaced between 133--129\,Ma, have any
extrusive equivalents in the Paran\'a LIP \citep{Peate.AGU-GM.97}. The dykes
are oriented orthogonal to our modeled initial extension direction, both
along the Brazilian as well as along the African margin in southern Angola.
Along with a predominant NE--SW striking metamorphic basement grain
presenting inherited weaknesses \citep{Almeida.ESR.00}, these dykes are
probably related to lithospheric extension along the conjugate southern
Campos/Santos--Benguela margin segment of the SARS.

In the southern segment, the Falkland-Aghulas Fracture zone is established
when the Maurice Ewing block starts moving with the Patagonian plate around
134\,Ma. Strain rate estimates from the Orange basin support
the onset of rifting around 150-145\,Ma \citep{Jones.G3.04}.

Towards the late Hauterivian (127\,Ma), the width of the enigmatic Pre-salt
Sag basin width has been created by slow, relative extension between SAm and
SAf, with extension velocities ranging between 7--9\,mm\,a$^{-1}$
(Campos/Jatob\'a, Fig.~\ref{fig:velplot}).

\subsection{Phase II: Equatorial Atlantic rupture (126.57--120.6\,Ma)}
\label{sub:intermediate_phase_equatorial_atlantic_rupture_126_57_120_6_ma_}

\begin{figure*}[t]
\vspace*{2mm}
\begin{center}
	\includegraphics[width=12cm]{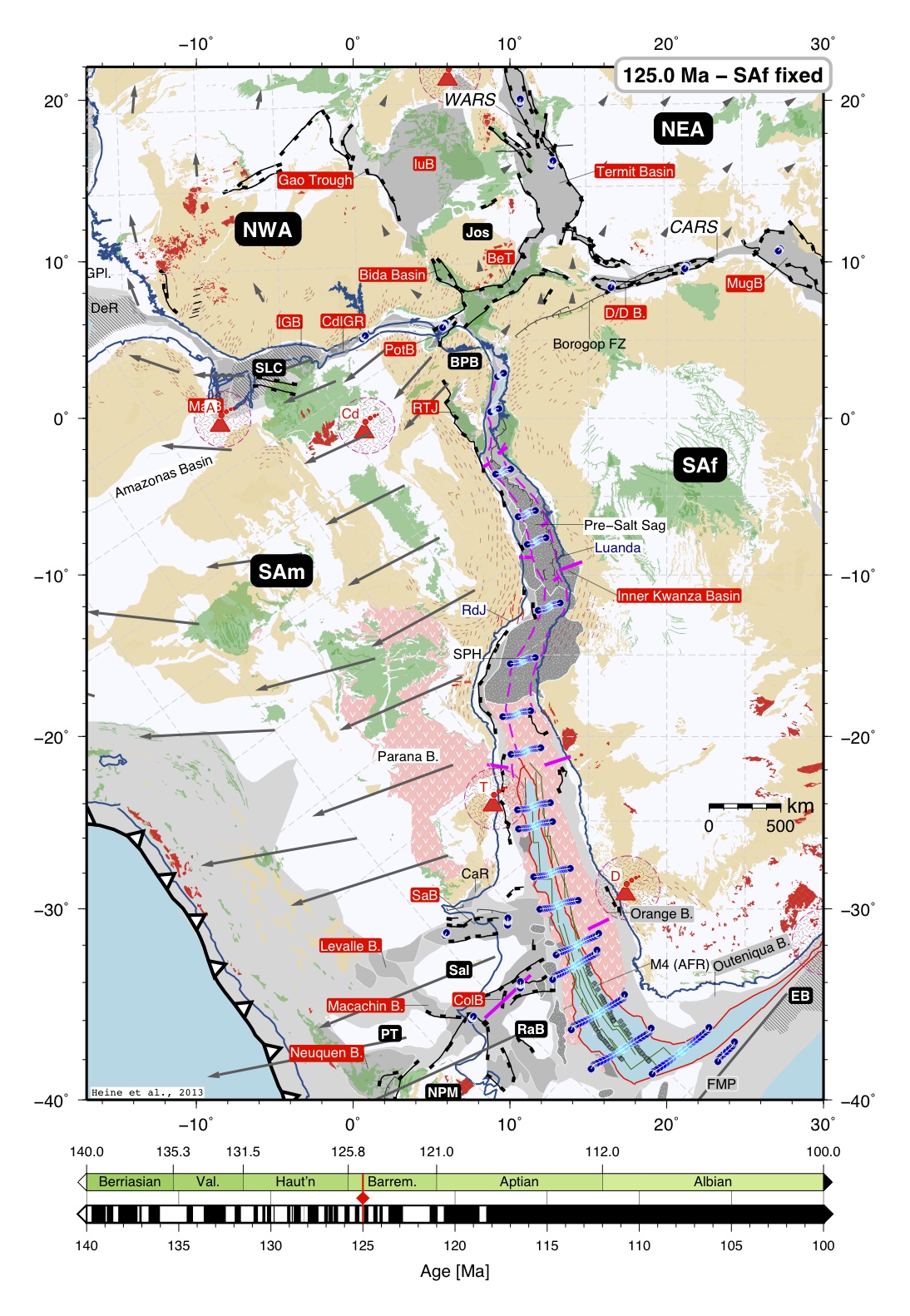}	
	\end{center}
		\caption[Reconstruction at 125\,Ma]{
		Plate tectonic reconstruction at 125\,Ma, with Africa fixed in present-day coordinates. For map legend see Fig.~\ref{fig:maplegend}, abbreviations as in Fig.~\ref{fig:140}. Note that at this time, the West African pre-salt basin width is generated and seafloor spreading is abutting against the Walvis ridge/Florianopolis ridge at the southern margin of the Santos Basin, with extension and possible rifting along the Abimael ridge in the inner SW part of the Santos basin \citep[cf.][]{Scotchman.GSL-PGC.10}. Relative rotation of SAm to NWA results in $\approx$20 km of transpression between the Guinea Plateau and the Demerara Rise. Rift basins along the Equatorial Atlantic rift are all actively subsiding. Additional abbreviations: SPH - S\~ao Paulo High, EB - Maurice Ewing Bank.
		\label{fig:125}
		}  
\end{figure*}

Following the initial rifting phase which is characterised by low strain
rates, deformation along the EqRS between NWA and northern SAm intensifies
due to strain localisation and lithospheric weakening
\citep{Heine.FragileEarth.11}. Marginal basins of the EqRS record increasing
rates of subsidence/transpression \citep{Azevedo.PhD.91} and in the southern
South Atlantic seafloor spreading anomalies M4 and M0 indicate a 3-fold
increase in relative plate velocities between SAm and African Plates
(Figs.~\ref{fig:125} and \ref{fig:velplot}). Velocity increases of similar
magnitude, albeit with a different timing, are reported by
\citet{Torsvik.GJI.09, Nuernberg.Tectp.91}. Along the EqRs,
\citet{Azevedo.PhD.91} describes a set of transpressional structures in the
Barreirinhas Basin which caused folding of Albian and older strata and
reports 50--120\,km of dextral strike slip along the Sobradinho Fault in the
proximal Barrerinhas basin, pointing towards an early dextral displacement.
With the changed kinematics, SAm now rotates clockwise around NWA, causing
transtensional opening of basins in the eastern part of the EqRS and about
20\,km of transpression along the Demerara Rise and Guinea Plateau at the
western end of the EqRS. This is in accordance with observed compressional
structures along the southern margin of the Guinea Plateau and the
northeastern margin of the Demerara Rise \citep{Basile.Tectp.13,
Benkhelil.Tectp.95}.

The increase in extensional velocities jointly occurs with a sudden change in
extension direction from NW--SE to more E--W (Fig.~\ref{fig:velplot}).
Depending on the distance from the stage pole for this time interval, the
directional change is between 75$^{\circ}$ in the northern SARS (Rio
Muni-Gabon/Potiguar-Sergipe Alagoas segment) and 30$^{\circ}$ in the southern
Pelotas/Walvis segment. The directional change visible in the flowlines
agrees very well with a pronounced outer gravity high along the Namibian
margin (Fig.~\ref{fig:flownamib}) and lineaments in the Santos Basin
(Fig.~\ref{fig:flowsantos}). This change in the plate motions had severe
effects on the patterns and distribution of extension in the SARS.
\citet{Karner.MPG.97} report a 100\,km westward step of the main axis of
lithospheric extension in the Gabon/Cabinda margin segment during their
second rift phase (Hauterivian to late Barr\^emian). Along the Rio Muni/
North Gabon margin a mild inversion in the pre-breakup sediment is observed
\citep{Lawrence.TLE.02} which we relate to the change in plate kinematics,
manifested in this margin segment as change from orthogonal/slightly oblique
extension to transform/strike slip. In the Santos Basin, the change in
extensional direction results in transtensional motions, most likely
responsible for the creation of the Cabo Frio fault zone and localised
thinning in proximal parts of the SW Santos basin, now manifested as an 
``Axis of Basement Low'' \citep{Modica.AB.04} north of the S\~ao Paulo High. By the time of change
in plate motions ($\approx$127\,Ma), the pre-salt sag  basin width had been fully
generated (Fig.~\ref{fig:125}). Increasing extensional velocities resulted in
fast, highly asymmetric localisation of lithosphere deformation in
the northern and central SARS segments, most likely additionally influenced by the
inherited basement grain/structures. This has left large parts of the Phase I
rift basin preserved along the Gabon-Kwanza margin, whereas south of the
Benguela/Cabo Frio transform the Phase I rift is largely preserved on the
Brazilian side.

\citet{Davison.GSLSP.07} reports that from about 124\,Ma \citep[Early Aptian
using][]{Gradstein.04}, extensive evaporite sequences are deposited in areas
north of the Walvis Ridge, starting with the Paripuiera salt in the northern
Gabon/Jequitinhonha segment and the slightly younger Loeme
evaporites in the Kwanza
basin. We note that in our reconstructions, the massive volcanic build ups of the Walvis Ridge/
Florianopolis Platform form a large barrier toward the Santos basin in the north.
North of the ridge, exension localised and propagated
northwards into the southwestern part of the Santos Basin \citep[the now
aborted Abimael ridge;][]{Scotchman.GSL-PGC.10, Mohriak.PG.10,
Gomes.AAPG-Conf.09} towards the end of this phase.

 The S\~ao Paulo High
remains part of the African lithosphere, which explains the ``Cabo Frio
counterregional fault trend'' in the northern Santos Basin
\citep{Modica.AB.04}. Prior NW-SE extension, paired with additional thinning
and transform motions along the Cabo Frio fault system
\citep{Stanton.Tectp.10, Meisling.AAPG-B.01} could have formed a shallow
gateway through the inner parts of the Santos Basin, allowing the supply of
seawater into the isolated central SARS. Dynamic, plume-induced
topography \citep{Rudge.EPSL.08} from the extensive magmatic activity in the Paran\'a-Etendeka
Igneous province might have caused additional lithospheric buoycancy of the
extending lithosphere in the Santos basin.

Seafloor spreading is fully established in a Proto-South Atlantic ocean basin
(Pelotas-Walvis segment to Falkland/Aghulas fracture zone), while oceanic 
circulation in this segment was probably quite restricted 
as the Falkland/Malvinas--Maurice Ewing Bank continental promontory had not 
cleared the Southern African margin (Fig.~\ref{fig:120}).

\subsection{Phase III: Break up -- Aptian to late Albian (120.6--100\,Ma)}
\label{sub:aptian_to_cenomanian_120_6_95_ma_}

\begin{figure}[t]
\vspace*{2mm}
\begin{center}
	\includegraphics[width=8.3cm]{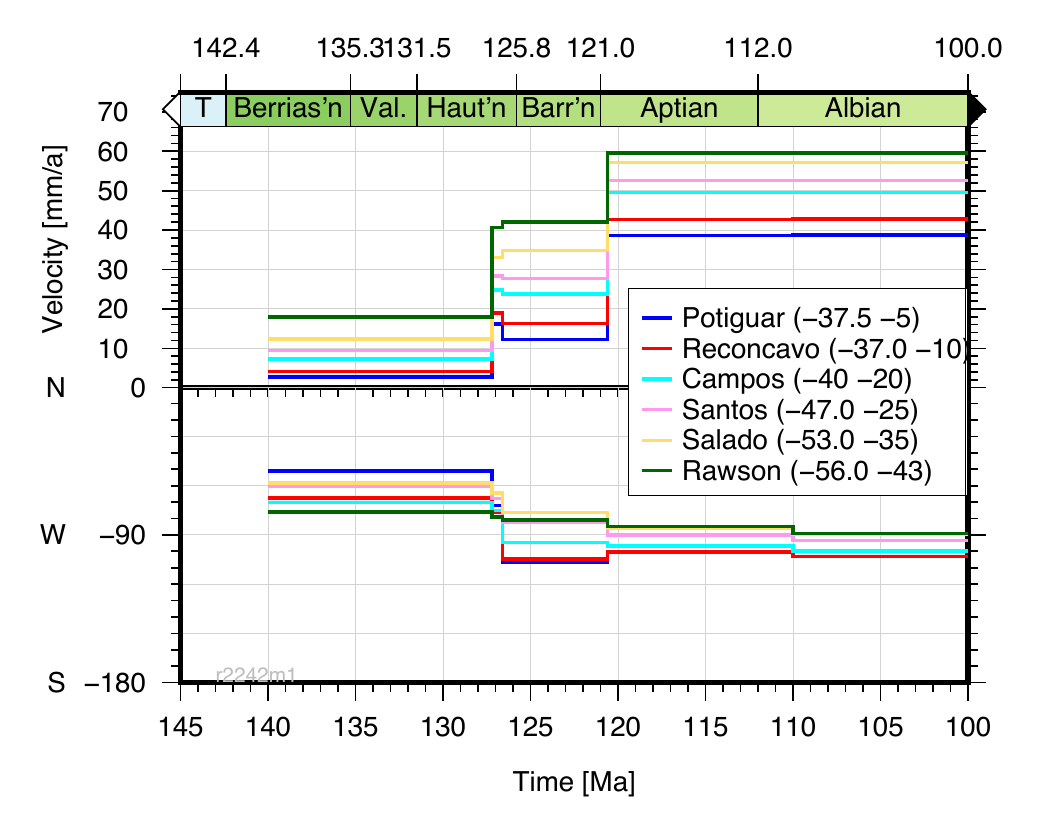}	
	\end{center}
		\caption{
		Relative separation velocities and directions between 6 South American and Patagonian points relative to a fixed Southern African plate using preferred plate kinematic model. Upper portion of figure shows extensional velocities over time, lower portion shows extension direction relative to a fixed Southern African plate.
		\label{fig:velplot}
	}
\end{figure}

Magnetic anomaly chron M0 (120.6\,Ma) is clearly identified in large parts of
the southern South Atlantic and provides, along with established fracture
zones, one major tiepoint for our reconstruction. Further lithospheric
weaking and strain localisation along the EqRS resulted in widespread
transtensional motions and related complex basin formation in the marginal
basins \citep{Basile.JAfES.05,Matos.AGU-GM.00,
DarrosDeMatos.GSLSP.99,Azevedo.PhD.91} causing a second increase in
extensional velocities between the African and SAm plates and a minor change
in separation direction from 120.6\,Ma onwards (Fig.~\ref{fig:velplot}). We
here linearly interpolate the plate velocities during the CNPS
(120.6--83.5\,Ma) only adjusting plate motion paths along well established
fracture zones.

\begin{figure*}[t]
\vspace*{2mm}
\begin{center}
	\includegraphics[width=12cm]{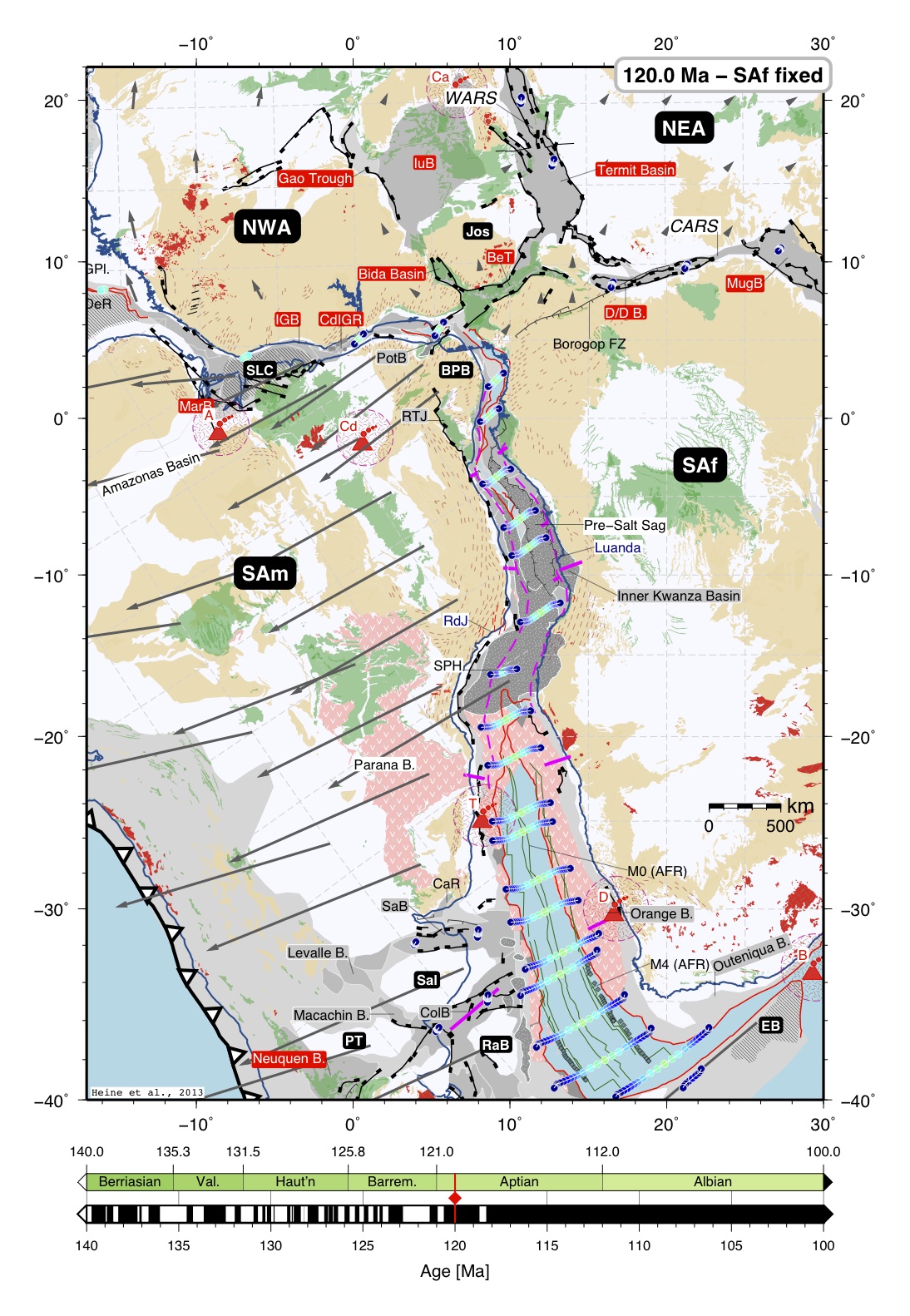}	
	\end{center}
		\caption[Reconstruction at 120\,Ma]{
		Plate tectonic reconstruction at 120\,Ma, with Africa fixed in present-day coordinates.For map legend see Fig.~\ref{fig:maplegend}, abbreviations as in Fig.~\ref{fig:140}. Significant increase in extensional velocities causes break up and subsequent seafloor spreading in the northernmost part of the South Atlantic rift extending down to the conjugate Cabinda/Espirito Santo segment. Due to changed kinematics, rifting and extension in the Santos/Benguela segment focusses on the African side, causing the transfer of the Sao Paulo High block onto the South American Plate. Breakup also occurs between Guinea Plateau and Demerara rise in the westernmost part of the Equatorial Atlantic.
		\label{fig:120}
		}  
\end{figure*}

By 120\,Ma (early Aptian, Fig.~\ref{fig:120}) continental breakup has
occurred in the northernmost SARS between the Potiguar/Benin and
Pernambuco--Paraiba/Rio Muni margin segments and between the conjugate
northern Gabon/Jequitinhonha--Camamu margins with incipient breakup in the
remaining part of the central SARS (Espirito Santo/Cabinda-Campos/Kwanza).
South of the Cabo Frio/Benguela transform, changes in plate motions resulted
in deformation to jump from the Avedis and Abimael ridges in the southwestern
Santos basin \citep{Scotchman.GSL-PGC.10, Mohriak.PG.10, Gomes.AAPG-Conf.09}
towards the African side, rifting the S\~ao Paulo High away from the
Namibian/Benguela margin. Extensive evaporite deposition continues in most
parts of the deforming central SARS, peaking in Mid-Aptian times
\citep{Karner.GSLSP.07, Davison.GSLSP.07}.

\begin{figure*}[t]
\vspace*{2mm}
\begin{center}
	\includegraphics[width=12cm]{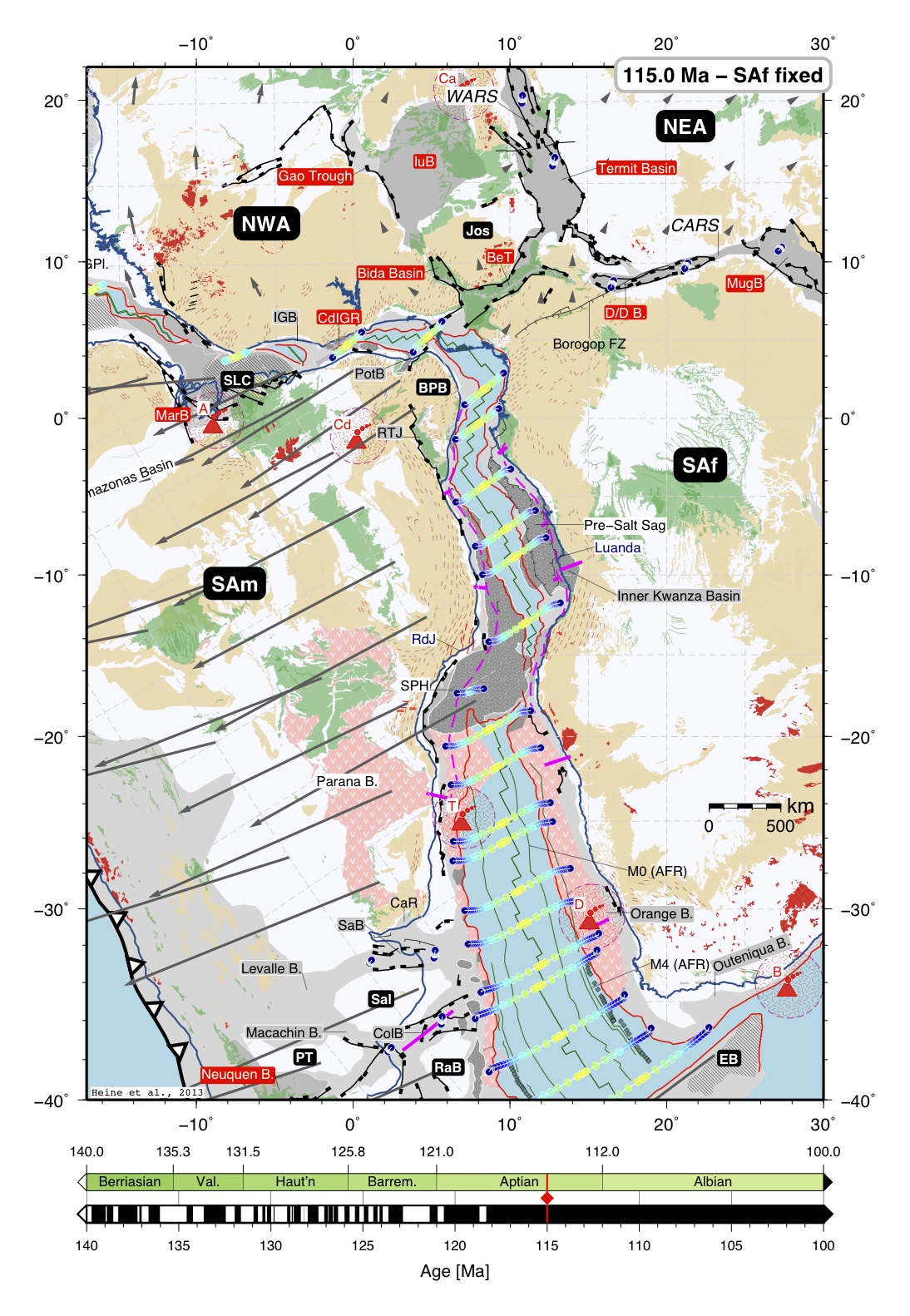}	
	\end{center}
		\caption[Reconstruction at 115\,Ma]{
		Plate tectonic reconstruction at 115\,Ma, with Africa fixed in present-day coordinates. For map legend see Fig.~\ref{fig:maplegend}, abbreviations as in Fig.~\ref{fig:140}. Continental break-up has occurred along most parts of the conjugate South American/Southern African and NW African margins apart from the ``Amazon'' and C\^ote d'Ivoire/Ghana transform segment of the Equatorial Atlantic. Oceanic accrection has already commenced in the Deep Ghanaian Basin. In the SARS, only the  Benguela-Walvis/Santos segment have not yet broken up. In this segment, extension is focussed asymmetrically close to the African margin, east of the S\~ao Paulo High. Basins in southern South America have entered the post rift phase and no significant relative motions between the rigid lithospheric blocks occur in post-Barr\^emian times. Post-rift thermal subsidence and possible gravitationally-induced flow of evaporite deposits towards the basin axis occurs in conjugate margin segments of the central SARS.
		\label{fig:115}
		}  
\end{figure*}

By 115\,Ma (Middle Aptian, Fig.~\ref{fig:115}), large parts of the SARS and
EqRS have broken up diachronously and entered post-rift thermal subsidence.
Our model predicts breakup in the Campos/Kwanza segment by 119\,Ma, in the
westernmost part of the EqRS at 118\,Ma, in line with deep water basin
conditions reported between the Guinea and St. Paul Fracture zones
\citep{Jones.G.87}. In the EqRS our model predicts accretion of oceanic
lithosphere in the Deep Ghanaian Basin from 117\,Ma onwards, and break up
along the southern Campos/Benguela margin segment in the SARS around
115\,Ma. For the Gabon margins, \citet{Dupre.MPG.07} and
\citet{Karner.MPG.97} assume the onset of rifting at around 118\,Ma, and Late
Barr\^emian--Early Aptian, respectively, which is in agreement with our
model. Subsidence data from nearly all margin segments in the SARS indicate
cessation of fault activity and a change to post-rift thermal subsidence by
this time, with the exception of the outer Santos Basin, where the final
breakup between SAf and SAm occurs between 113--112\,Ma. This timing is in
agreement with the deposition of the youngest evaporites in the outer Santos
Basin around 113\,Ma \citep{Davison.GSLSP.07}. We hypothesise that the early
salt movement in the Gabon, Kwanza, Espirito Santo, Campos and Santos basins
and the observed chaotic salt in the distal part of these basins is related
to the fast localisation of lithospheric deformation, break-up and subsequent
rapid subsidence during the early Aptian, introducing topographic gradients
favouring gravitational sliding and downslope compression in the earliest
postrift \citep{Fort.AAPG-B.04}.

\begin{figure*}[t]
\vspace*{2mm}
\begin{center}
    \includegraphics[width=12cm]{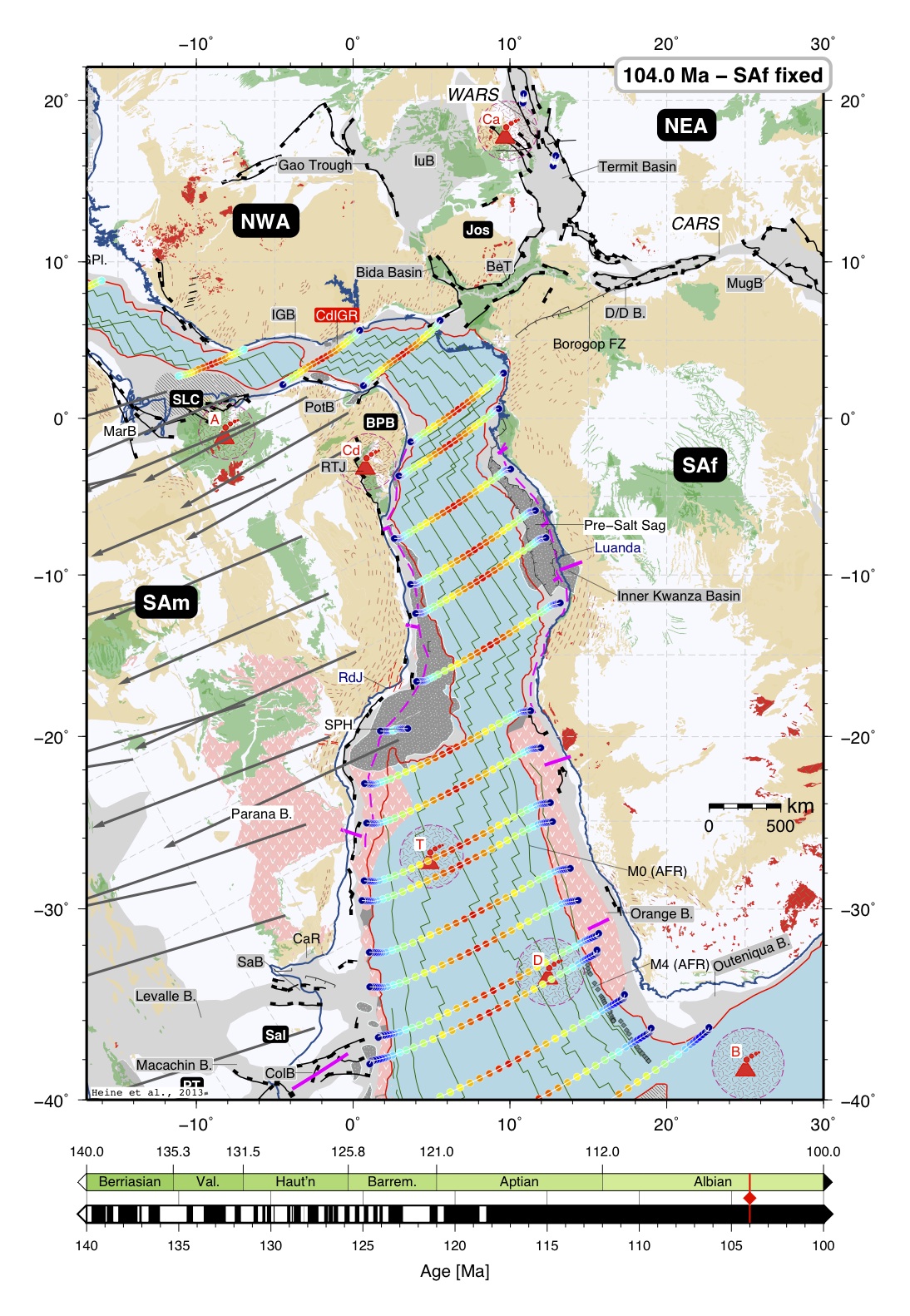}
    \end{center}
        \caption[Reconstruction at 104\,Ma]{
        Plate tectonic reconstruction at 104\,Ma, with Africa fixed in present-day coordinates. For map legend see Fig.~\ref{fig:maplegend}, abbreviations as in Fig.~\ref{fig:140}. Full continental separation is achieved at this time, with narrow oceanic gateways now opening between the C\^ote d'Ivoire/Ghana Ridge and the Piaui-C\'eara margin in the proto-Equatorial Atlantic and between the Ewing Bank and Aghulas Arch in the southernmost South Atlantic. Deformation related to the break up between Africa and South America in the African intracontinental rifts ceases in post-Aptian times.
        \label{fig:104}
        }  
\end{figure*}

Relative plate motions in the CARS and WARS cease around 110\,Ma in the early
Albian, with most intracontinental rift basins entering a phase of thermal
subsidence before subsequent minor reactivation occurred in Post-Early
Cretaceous times \citep{Janssen.GSA-B.95,Genik.AAPG-B.93,Maurin.Tectp.93,
Genik.Tectp.92, Guiraud.Tectp.92a, McHargue.Tectp.92}. The cessation of
rifting in WARS and CARS results in the onset of transpression along the
C\^ote d'Ivoire-Ghanaian Ridge in the EqRS as the trailing edges of SAm
continue to move westward, while NWA now remains stationary with respect to the
other African plates. The NWA transform margins now provide a backstop to
the westward-directed motions of SAm, causing compression associated with up
to 2\,km of uplift to occur along the Ghanaian transform margin between the
middle to late Albian \citep{Antobreh.MPG.09, Basile.JAfES.05,
Pletsch.JSAES.01, Clift.JGSL.97}. Complete separation between African and
South American continental lithospheres is achieved at 104\,Ma
(Fig.~\ref{fig:104}), while the oceanic spreading ridge clears the C\^ote
d'Ivoire/Ghana Ridge by 100--99\,Ma.

\conclusions \label{sec:conclusions}

We present a new plate kinematic model for the evolution of the South
Atlantic rifts. Our model integrates intraplate deformation from temporary
plate boundary zones along the West African and Central African Rift Systems
as well as from Late Jurassic/Early Cretaceous rift basins in South America
to achieve a tight fit reconstruction between the major plates. Our plate motion hierarchy 
describes the motions of South America to
Southern Africa through relative plate motions between African sub-plates, allowing to 
to model the time-dependent pre-breakup extension history of the
South Atlantic rift system. Three main phases with distinct velocity and
kinematics result through the extension along the Central African, West
African and Equatorial Atlantic rift systems exerting significant control on the
dynamics of continental lithospheric extension in the evolving South Atlantic
rift from the Early Cretaceous to final separation of Africa and South
America in the late Albian (104\,Ma). An intial phase of slow E--W extension in the South Atlantic rift basin from
140\,Ma (Base Cretaceous) to 127\,Ma (late Hauterivian) causes distributed
extension in W--E direction. The second phase from 126--121\,Ma (late
Hauterivian to base Aptian) is characterised by rapid lithospheric weakening
along the Equatorial Atlantic rift resulting in increased extensional
velocities and a change in extension direction to SW--NE. In the last phase commences at 120\,Ma
and culminates in diachronous breakup along the South Atlantic rift and
formation of the South Atlantic ocean basin. It is
characterised by a further increase in plate velocities with a minor change
in extension direction.

We argue that our proposed three-stage kinematic history can account for most
basin forming events in the Early Cretaceous South Atlantic, Equatorial
Atlantic, Central African and West African Rift Systems and provide a robust
quantitative tectonic framework for the formation of the conjugate South
Atlantic margins. In particular, our model addresses the following issues
related to the South Atlantic basin formation:
\begin{itemize}
\item We achieve a tight fit reconstruction based on structural restoration of the
conjugate South Atlantic margins and intraplate rifts, without the need for
complex intracontinental shear zones.
\item The orientation of the Colorado and Salado Basins, oriented perpendicular to
the main South Atlantic rift are explained through clockwise rotation of the
Patagonian sub-plates. This also implies that the southernmost South Atlantic
rift opened obliquely in a NE--SW direction.
\item A rigid South American plate, spanning the Camamu to Pelotas basin margin segments,
requires that rifting started contemporaneously in this segment of the South Atlantic Rift
(Gabon, Kwanza, Benguela and Camamu, Espirito Santo, Campos and Santos basins), albeit so far stratigraphic 
data \citep[e.g.][]{Chaboureau.Tectp.12} indicates differential onset of the syn rift phase
in the various marginal basins.
\item The formation of the enigmatic pre-salt sag  basin along the West African margin
is explained through a multi-phase and multi-directional extension history in
which the initial sag basin width was created through slow (7--9\,mm\,a$^{-1}$)
continental lithospheric extension until 126\,Ma (Late Hauterivian).
\item Normal seafloor spreading in the southern segment of the South Atlantic rift commenced
at around 127\,Ma, after a prolonged phase of volcanism affecting the southern South Atlantic conjugate margins.
\item Strain weakening along the Equatorial Atlantic Rift caused an at least 3-fold increase
of plate velocities between South America and Africa between 126--121\,Ma,
resulting in rapid localisation of extension along the central South Atlantic
rift.
\item Linear interpolation of plate motions between 120.6\,Ma and 83.5\,Ma yield break-up ages
for the conjugate Brazilian and West African margins which are corroborated
by the onset of post-rift subsidence in those basins.
\end{itemize}

We conclude that our multi-direction, multi-velocity extension history can consistently
explain the key events related to continental breakup in the South Atlantic
rift realm, including the formation of the enigmatic pre-salt sag basin. 

The plate kinematic framework presented in this paper is robust and comprehensive, 
shedding new light on the spatio-temporal
evolution of the evolving South Atlantic rift system. However, problems remain with regard
to the absolute timing of events in the evolution of the South Atlantic rift due to a lacking
coherent stratigraphy for both conjugate margin systems, and limited access to crustal scale
seismic data. Relatively old and limited data for the African intraplate rifts results in large uncertainties
in both absolute timing as well as structural constraints on the kinematics of rifting which we have attempted to minimise in this comprehensive approach.

Strengthening the kinematic framework for the South Atlantic rift also offers an ideal 
laboratory to investigate the interaction between large scale plate tectonics and 
lithosphere dynamics during rifting, and the relationship between plume activity,
rift kinematics and  the formation of volcanic/non-volcanic margins.

Data and high resolution images related to the publication are provided with the manuscript and 
can be freely accessed under a Creative Commons license at the Datahub  under the following URL:
\url{http://datahub.io/en/dataset/southatlanticrift}. 

\begin{acknowledgements}

We would like to thank Statoil for permission to publish this work. CH and JZ
are indebted to many colleagues in Statoil's South Atlantic new venture
teams, specialist groups, and R\&D who have contributed in various aspects to
an inital version of the model. In particular we would like to thank Hans
Christian Briseid, Leo Duerto, Rob Hunsdale, Rosie Fletcher, Eric Blanc,
Tommy Mogensen Egebjerg, Erling V{\aa}gnes, Jakob Skogseid, and Ana
Serrano-O\~nate for discussions and valuable feedback. Feedback from Patrick
Unternehr, Ritske Huismans, Julia Autin and Hannes Koopmann on
various aspects of our work are kindly acknowledged. CH thanks
John Cannon and Paul Wessel for numerous fast software bug fixes. 
Editor Douwe J. J. v. Hinsbergen, an anonymous reviewer and Dieter 
Franke are thanked for constructive reviews and comments which improved 
the discussion version of the paper. Dieter Franke also kindly 
supplied digitised SDR outlines and structural data for the Argentine margin.

Plate reconstructions were made using the open-source software GPlates
(\url{http://www.gplates.org}), all figures were generated using GMT
\citep[\url{http://gmt.soest.hawaii.edu};][]{Wessel.EOS.98}. Structural
restorations were partly done using Midland Valley's Move software with an
 academic software license to The University of Sydney.
 
 CH is funded by ARC Linkage Project L1759 supported by Shell International E\&P and
 TOTAL. JZ publishes with permission from Statoil ASA. RDM is supported by ARC
 Laureate Fellowship FL0992245. 
 
\end{acknowledgements}

%------------------------------------------------------------------------------
%-- BIBLIOGRAPHY
%% Literature citations
%% command                        & example result
%% \citet{jones90}|               & Jones et al.\ (1990)
%% \citep{jones90}|               & (Jones et al., 1990)
%% \citep{jones90,jones93}|       & (Jones et al., 1990, 1993)
%% \citep[p.~32]{jones90}|        & (Jones et al., 1990, p.~32)
%% \citep[e.g.,][]{jones90}|      & (e.g., Jones et al., 1990)
%% \citep[e.g.,][p.~32]{jones90}| & (e.g., Jones et al., 1990, p.~32)
%% \citeauthor{jones90}|          & Jones et al.
%% \citeyear{jones90}|            & 1990

\bibliographystyle{copernicus}
\bibliography{HeineZouthout_SthAtlanticRiftKinematics}

%----------------------------------------------------------------------------%
%:==> FIGURES
%----------------------------------------------------------------------------%

%---- FIGURES AT THE END OF SUBMISSION
%% ONE-COLUMN FIGURES
%f
% \begin{figure}[t]
% \vspace*{2mm}
% \begin{center}
% \includegraphics[width=8.3cm]{FILE NAME}
% \end{center}
% \caption{TEXT}
% \end{figure}

% %% TWO-COLUMN FIGURES
% %f
% \begin{figure*}[t]
% \vspace*{2mm}
% \begin{center}
% \includegraphics[width=12cm]{FILE NAME}
% \end{center}
% \caption{TEXT}
% \end{figure*}

% -----------------------------------------------------------------------------
%: FIG 1 Overview - Rigid plates and deforming zones
\clearpage

% -----------------------------------------------------------------------------
%: FIG:2 Time scale comparison

\clearpage

% -----------------------------------------------------------------------------
%: FIG 3: Restored margin width based on cross sections

\clearpage

% -----------------------------------------------------------------------------
%: FIG 4 -- Africa overview
\clearpage

% -----------------------------------------------------------------------------
%: FIG 5 - Extension estimates Termit

\clearpage

% -----------------------------------------------------------------------------
%: FIG 5 - Extension estimates CARS

\clearpage

% -----------------------------------------------------------------------------
%: FIG 6 - Extension estimates WARS

\clearpage

% -----------------------------------------------------------------------------
%: FIG 7 -- SAM Overview
\clearpage

% -----------------------------------------------------------------------------
%: FIG 8 -- Plate circuit

\clearpage

% -----------------------------------------------------------------------------
%: FIG 9 -- Flowline Gabon
\clearpage
% -----------------------------------------------------------------------------
%: FIG 10 -- Flowline Orange
\clearpage

% -----------------------------------------------------------------------------
%: FIG 11 -- Flowline Santos
\clearpage
 
% -----------------------------------------------------------------------------
%: FIG 12 reconstruction at 143 ma

% -----------------------------------------------------------------------------
%: FIG 14 -- Overlap

\clearpage

% -----------------------------------------------------------------------------
%: FIG 15 -- Velocities

\clearpage
 
% -----------------------------------------------------------------------------
%: FIG 16 -- reconstruction at 132 ma

% -----------------------------------------------------------------------------
%: FIG 17 -- reconstruction at 125 ma

% -----------------------------------------------------------------------------
%: FIG 18 -- reconstruction at 120 ma

% -----------------------------------------------------------------------------
%: FIG 19 -- reconstruction at 115 ma

% -----------------------------------------------------------------------------
%: FIG 20 -- reconstruction at 110 ma

% -----------------------------------------------------------------------------
% : FIG 21 -- reconstruction at 104 ma

% -----------------------------------------------------------------------------
%: FIG 22 -- map legend

\begin{figure*}[t]
\vspace*{2mm}
\begin{center}
	\includegraphics[width=12cm]{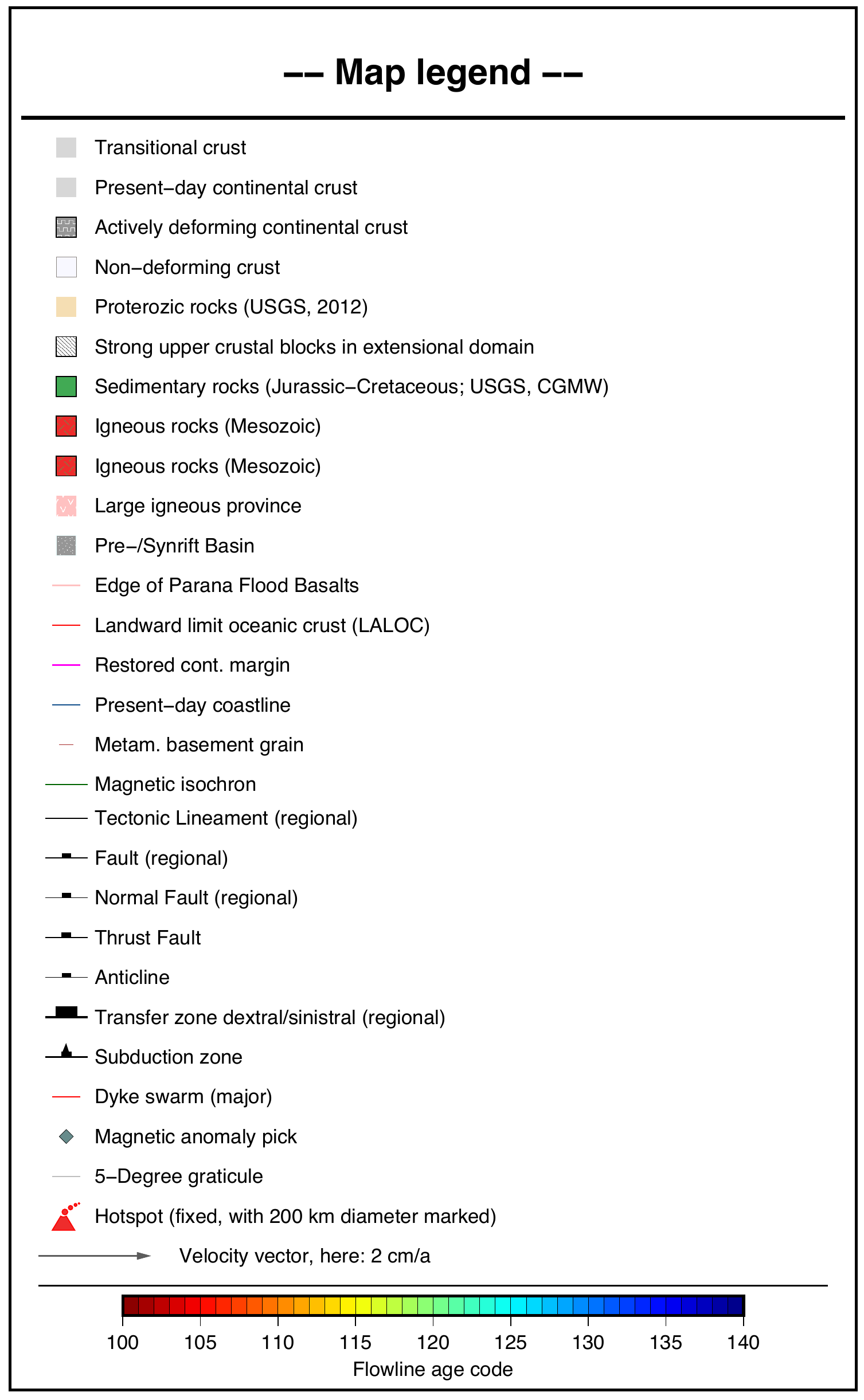}	
	\end{center}
		\caption[Map legend for reconstructions]{
		Map legend for Figs.~\ref{fig:140}-\ref{fig:104}.
		\label{fig:maplegend}
		}  
\end{figure*}

\clearpage

% -----------------------------------------------------------------------------
%:==> TABLES

\begin{table*}[t]
\caption{Finite rotations parameters for main rigid plates for preferred South Atlantic rift model M1. Abbreviations are Abs - absolute reference frame, SAf - Southern Africa, NEA - Northeast Africa, NWA - Northwest Africa, Sal - Salado Subplate, NPM - North Patagonian massif block, CM - Mesozoic magnetic anomaly chron, y - young, o - old. magnetic anomaly timescale after \citet{Gee.ToG.07}. For microplate rotation parameters please see electronic supplements.
\label{tab:rotationparams}
}
\vskip4mm
\centering
\begin{tabular}{lllllll}
\tophline
Moving  &   & \multicolumn{3}{c}{Rotation pole} & Fixed  & Comment\\
plate   &   Age   & Lon & Lat & Angle                 & plate  & \\
\middlehline
SAf & 100.00 & 14.40 & -29.63 & -20.08 & Abs & \citet{ONeill.G3.05} \\
SAf & 110.00 &  6.61 & -29.50 & -26.77 & Abs & \citet{Steinberger.Nat.08}\\
SAf & 120.00 &  6.11 & -25.08 & -30.45 & Abs & \citet{Steinberger.Nat.08}\\
SAf & 130.00 &  5.89 & -25.36 & -33.75 & Abs & \citet{Steinberger.Nat.08}\\
SAf & 140.00 &  7.58 & -25.91 & -38.53 & Abs & \citet{Steinberger.Nat.08}\\
SAf & 150.00 & 10.31 & -27.71 & -37.25 & Abs & \citet{Steinberger.Nat.08}\\
\middlehline
NEA & 110.0  &  0.0  &  0.0   &   0.0  & SAf & \\
NEA & 140.0  &  -0.13  & 39.91   &  1.35  & SAf & Fit\\ 
\middlehline
NWA & 110.0  &  0.0  &  0.0   &   0.0  & NEA & \\
NWA & 140.0  & 25.21 & 5.47   &  2.87  & NEA & Fit\\
\middlehline
SAm & 83.5   & 61.88 & -34.26 & 33.51  & SAf & \citet{Nuernberg.Tectp.91}, CAn34\\
SAm & 96.0   & 57.46 & -34.02 & 39.79  & SAf & Linear interpolation\\
SAm & 120.6  & 51.28 & -33.67 & 52.35  & SAf & Anomaly CM0ry\\
SAm & 120.6  & 52.26 & -34.83 & 51.48  & NWA & Crossover CM0ry \\
SAm & 126.57 &  50.91 &  -34.59 &  52.92  & NWA & Anomaly CM4no\\
SAm & 127.23 &  50.78  &  -34.54 &  53.04  & NWA & Anomaly CM7ny\\
SAm & 140.0  & 50.44 & -34.38 & 53.40  & NWA & Fit\\
\middlehline
NPM & 124.05 &   0.0  &   0.0  &  0.0  & Sal & \\
NPM & 150.00 & -35.94 & -63.06 & 4.10  & Sal & Fit - Extension based on \citet{Pangaro.MPG.12}\\
\middlehline
Sal & 124.05 &   0.0  &  0.0   & 0.0   & SAm & \\
Sal & 145.00 & -33.02 & -60.52 & 4.40  & SAm & Fit\\ 
\bottomhline
\end{tabular}
\end{table*}

%-------------------

\begin{table*}[t]
\caption{Alternative kinematic scenarios tested for onset and cessation of syn-rift phases in main rift zones.\label{tab:kinmodels}}
\vskip4mm
\centering
\begin{tabular}{lp{8cm}l}
\tophline
Model & Syn-rift onset  & Syn-rift cessation in WARS and CARS\\
\middlehline
PM1 & \textbf{SARS, CARS, WARS, and EqRS simultaneously}: around Base Cretaceous (140\,Ma). & 110\,Ma (early Albian)\\
PM2	& \textbf{SARS, CARS, WARS, and EqRS simultaneously}: around Base Cretaceous (140\,Ma). & 100\,Ma (late Albian)\\
PM3	& \textbf{SARS, CARS and WARS simultaneously}:  around Base Cretaceous (140\,Ma); \textbf{EqRS}: Top Valangian (132\,Ma)& 110\,Ma (early Albian)\\
PM4	& \textbf{SARS, EqRS, and CARS simultaneously}: around Base Cretaceous (140\,Ma); \textbf{WARS}: Base Valangian (135\,Ma) & 110\,Ma (early Albian)\\
PM5	& \textbf{SARS and EqRS} simultaneously around Base Cretaceous (140\,Ma); \textbf{WARS and CARS}: 135\,Ma & 110\,Ma (early Albian)\\
NT91 & \multicolumn{2}{l}{Model parameters of \citet{Nuernberg.Tectp.91} with forced breakup at 112\,Ma after \citet{Torsvik.GJI.09}} \\
\bottomhline
\end{tabular}
\end{table*}

%-------------------

\begin{table*}[t]
\caption{Deformation estimates used for major intraplate deformation zones. ``Computed'' extension is based on the our method as described in Sec.~\ref{sec:data_and_meth} and represents a maximum estimate as it is based on total sediment thickness, always measured orthogonal to major basin/rift axis. ``Implemented'' deformation is actual displacement as implemented in plate kinematic model. Integers in column ``Deforming zone'' refer to plate circuit (Fig.~\ref{fig:platecircuit}), abbreviations refer to plate pairs. $\perp$: extension measured orthogonal to long rift/basin axis, sc: along small circle around corresponding stage pole.\label{tab:extnestimates}}
\vskip4mm
\centering
\begin{tabular}{llp{2cm}p{2cm}p{2cm}l}
\textbf{Deforming} & \textbf{Basin} & \multicolumn{3}{c}{\textbf{Extension} [km]}            & \textbf{Syn-rift duration} \\
\textbf{zone}      & \textbf{names}               & Published & Computed & Implemented       & [Myrs] \\
\tophline
\multicolumn{6}{l}{\textbf{CARS - Central African Rift System}}\\ % CARS\nMuglad/Melut\nSalamat/Doseo Basins
1 - NEA-SAf & Melut Basin &  15\,km$^a$ & 76\,km (max.) & see Muglad & 140--110\,Ma\\
1 - NEA-SAf & Muglad Basin & 24--27\,km$^a$,22--48\,km$^b$, 56$\pm6$$^c$\,km & 83\,km (max.) & 50\,km (N), 27\,km (S) & 140--110\,Ma\\
1 - NEA-SAf & Doseo Basin & 35-40\,km$^d$ (dextral) & 41\,km (Doseo) $\perp$  & 52\,km (oblique), 30\,km (E-W) & 140--110\,Ma\\
2 - BON-NEA & Bongor/Doba Basins &  & 31\,km (Bongor, $\perp$), 20\,km (Doba, $\perp$) & 57\,km for Doba \& Bongor basins comb. (sc) & 140--100\,Ma\\
\middlehline
\multicolumn{6}{l}{\textbf{WARS - West African Rift System}}\\ % WARS\nTermit,Tenere,\nGrein-Kafra
3 NWA-NEA & Termit Basin & 40\,km$^d$ & 27--98\,km & 60-65\,km $\perp$ , 80-85\,km (sc) & 140--110\,Ma\\
% 2 NWA-NEA & T\'en\'er\'e Rift  & 70 km & & & 33\\
3 NWA-NEA & Grein-Kafra Basin &  (?) & 61\,km (max.) & 22\,km $\perp$, 60-65\,km (sc) & 140--110\,Ma\\
4 JOS-NWA & Bida/Gao/Iullemmeden  & (?) & 28\,km (Bida/Iul\-lem\-meden, max.) & 15-20\,km (Gao Trough, sc) & 135--110\,Ma\\ 
JOS $|$ BEN & (Central) Benoue Trough  &  & 61\,km (max.) & 36\,km $\perp$, 40\,km (sc) & 140--110\,Ma\\ 
\middlehline
\multicolumn{6}{l}{\textbf{South America}}\\
5 - SAL-SAm & Salado/Punta Del Este Basins & (?) & 65\,km (max.) & 45\,km (sc, $\perp$) & 140--120.6\,Ma\\
6 - NPM-SAL & Colorado Basin & 45\,km$^b$ & 106\,km (max.) & 40\,km (sc, $\perp$) & 140--120.6\,Ma\\
\bottomhline
\multicolumn{6}{l}{$^a$\citet{McHargue.Tectp.92}, $^b$\citet{Browne.Tectp.83}, $^c$\citet{Mohamed.JAfES.01},$^d$\citet{Genik.Tectp.92},$^e$\citet{Pangaro.MPG.12}}\\
% \multicolumn{6}{l}{}\\
% \multicolumn{6}{l}{}\\

\end{tabular}
\end{table*}

%% The different columns must be seperated with a & command and should
%% end with \\ to identify the column brake.

%%%%%%%%%%%%%%%%%%%%%%%%%%%%%%%%%%%%%%%%%%%%%%%%%%%%%%%%%%%%%%%%%%%%%%%%%%%%%%

%% If figures and tables must be numbered 1a, 1b, etc. the following command
%% should be inserted before the begin{} command.

% \addtocounter{figure}{-1}\renewcommand{\thefigure}{\arabic{figure}a}

\end{document}